\documentclass[11pt]
{article}
\usepackage{amsmath}
\usepackage{amsfonts}
\usepackage{amssymb}
\usepackage{epsfig}

\DeclareMathOperator{\supp}{supp}

\newcommand{\dd}{\mathrm{d}}
\newcommand{\ddt}{\frac{\partial\phantom{t}}{\partial t}}
\newcommand{\ddtt}{\frac{\partial^2\phantom{t}}{\partial t^2}}

\newcommand{\eb}{e_{\text{b}}}

\newcommand{\mb}{m_{\text{b}}}

\newcommand{\mbrot}{m_{\text{b},\omega}}
\newcommand{\mfrot}{m_{\text{f},\omega}}

\newcommand{\Ib}{I_{\text{b}}}
\newcommand{\Ibrot}{I_{\text{b},\omega}}
\newcommand{\Ifrot}{I_{\text{f},\omega}}

\newcommand{\emp}{_{\text{e}}}

\newcommand{\Id}{{\mathrm{I}}}

\newcommand{\fe}{f_{\text{e}}}
\newcommand{\fm}{f_{\text{m}}}

\newcommand{\vect}[1] {\vec{{ #1}} }

\newcommand{\pkc}{p_{\text{c},k}}	

\newcommand{\pkb}{p_{\text{b},k}}	

\newcommand{\qE}{q_{\text{e}}}		
\newcommand{\omE}{\omega_{\text{e}}}	

\newcommand{\pb}{p_{\text{b}}}		
\newcommand{\sbE}{s_{\text{b}}}         
\newcommand{\mub}{\mu_{\text{b}}}       

\newcommand{\sbf}{s_{\text{bf}}}        
 
\newcommand{\gb}{g_{\text{b}}}          

\newcommand{\jV}{{\vect{j}}}		
\newcommand{\qV}{{\vect{q}}}            
\newcommand{\vV}{{\vect{v}}}            

\newcommand{\nab}{{\nabla}}
\newcommand{\nabE}{{\nab_{\mathrm{e}}}}

\newcommand{\veps}{{\varepsilon}}
\renewcommand{\leq}{\leqslant}
\renewcommand{\geq}{\geqslant}

\newcommand{\Cset}{\mathbb{C}}
\newcommand{\Rset}{\mathbb{R}}

\newcommand{\rhoel}{\rho_{\mathrm{el}}}
\newcommand{\jel}{j_{\mathrm{el}}}
\newcommand{\rhoelext}{\rho_{\mathrm{el}}^{\mathrm{ext}}}
\newcommand{\jelext}{j_{\mathrm{el}}^{\mathrm{ext}}}

\renewcommand{\Re}{\mathrm{Re\,}}

\newcommand{\half}{{\textstyle{\frac{1}{2}}}}

\begin{document}

\title{Quantum Abraham models with de Broglie-Bohm laws 
		of electron motion\footnote{This corrected
	preprint was typeset with \LaTeX. The original
	version (four typos) appeared in
			``Quantum Mechanics'' 
	(Festschrift for G.C. Ghirardi's 70th birthday), 
	A. Bassi, D. Duerr, T. Weber and N. Zanghi (Eds.), 
	AIP Conference Proceedings v. 844, pp. 206--227 (2006).
     Copyright for the Festschrift version has been transferred to the AIP.
	In this corrected version also an appendix has been added.}}

\author{M\sc{ichael} K.-H. K\sc{iessling}\\
Department of Mathematics, Rutgers University\\
	 110 Frelinghuysen Rd.,	Piscataway, N.J. 08854}

\maketitle

\begin{abstract}
\noindent
	We discuss a class of quantum Abraham models in which the $N$-particle spinor wave function of $N$ 
electrons solves a Pauli respectively Schr\"odinger equation, featuring regularized classical electromagnetic 
potentials which solve the semi-relativistic Maxwell-Lorentz equations for regularized 
point charges, which move according to some de Broglie-Bohm law of quantum motion. 
	Thus there is a feedback loop from the actual particle motions to the wave function.
	The electrons have a bare charge and positive bare mass different from
their empirical charge and mass due to renormalization by the self-fields.
	In the classical limit the various models reduce to the
Hamilton-Jacobi version of corresponding Abraham models of classical electron theory. 
\end{abstract}

%
\noindent
PACS \#: {03.65.Ta 	
		03.50.De 	
		}

\noindent
\small{KEYWORDS: Classical electromagnetic fields, Abraham model, de Broglie-Bohm type quantum theory}

\newpage

\section{Introduction}
	Beside GRW quantum mechanics \cite{GRW}, de Broglie-Bohm quantum mechanics
\cite{BohmsHIDDENvarPAPERS, dbBOOK, bohmhileyBOOK, bellBOOK}, a.k.a. Bohmian mechanics \cite{detlefBOOK}, 
is another ontologically clear theory. 
	Consider $N$ not mutually interacting nonrelativistic spin 1/2 electrons, each having charge $-e$ 
and empirical mass $m$. 
	The electrons are regarded as real particles 
with \textit{actual} positions $q_k(t)\in \Rset^3,\ k=1,...,N$ which, when grouped together as
$(q_1,...,q_N)(t) = \qV(t)\in \Rset^{3N}$, move according to the guiding equation
\begin{equation}
 \frac{\dd \qV}{\dd t} 
\Big|_{\qV=\qV(t)} = \vV(\qV(t),t),
\label{eq:dBBguide}
\end{equation}
where the velocity field $\vV$ on labeled $N$ particle configuration space $\Rset^{3N}$ is defined by
$\rho\vV=\jV$,
where $\rho = \psi^\dagger\psi$ is the conventional Born formula for the 
``probability'' density\footnote{In de Broglie-Bohm type
	quantum theory, the true nature of $\psi$ is dynamical and the probabilistic spinoff merely a 
	consequence of a suitable randomness in the initial positions and the equivariance of 
	$\psi^\dagger\psi$ (spinor space inner product).}
and $\jV$ is a probability current density vector associated to $\psi$ by the continuity
equation $\partial_t\rho +\vec{\nab}\cdot\jV =0$.
	Clearly, $\jV$  is only determined up to an additive curl.
	Two possibilities have been proposed in the literature,
see \cite{bohmhileyBOOK} and  \cite{bellBOOK}.
	Writing $\jV=  (j_1,...,j_N)$, the minimal choice suggested by the Pauli equation
leads to the  componentwise definition  \cite{bellBOOK}
\begin{equation} 
j_k
= 
 {\Re\,\psi^\dagger\left(-i{\textstyle{\frac{\hbar}{m}}}\nab_k + {\textstyle{\frac{e}{mc}}}A_k\right)\psi},
\phantom{+ \frac{\hbar}{2m}{\nab_k\times\left(\psi^\dagger\sigma_k\psi\right)}}
\label{eq:jA}
\end{equation}
while the non-relativistic limit of Dirac's current suggests \cite{bohmhileyBOOK},
\begin{equation} 
j_k
= 
{\Re\,\psi^\dagger\left(-i{\textstyle{\frac{\hbar}{m}}}\nab_k + {\textstyle{\frac{e}{mc}}}A_k\right)\psi}
+ {\textstyle{\frac{\hbar}{2m}}}{\nab_k\times\left(\psi^\dagger\sigma_k\psi\right)}.
\label{eq:jB}
\end{equation}
	In \eqref{eq:jB},
either $\sigma_k= (\sigma^x,\sigma^y,\sigma^z)_k$, where $\sigma^{x,y,z}$ are
the Pauli spin operators, or $\sigma_k= (\Id,\Id,\Id)_k$, where $\Id$ is the identity operator.
	In either case $\sigma_k$ acts on the $k$-th spin-up-down variable $\zeta_k\in\{-1,1\}$ 
of the electrons' antisymmetric $N$-particle spinor wave function 
$\psi(\,.\,,t) \in \Cset^{\Rset^{3N}\times\{-1,1\}^N}$, 
which is normalized as $\int_{\Rset^{3N}}(\psi^\dagger\psi)(q_1,...,q_N) \dd^{3N}q =1$
and solves the Pauli equation
\begin{equation}
i\hbar\ddt\psi 
= 
\sum_{k=1}^N \Big({\textstyle{\frac{1}{2m}}}
 \left( \sigma_k\cdot\left(-i\hbar\nab_k + \textstyle{\frac{e}{c}}A_k\right)\right)^2 -e\phi_k\Big)\psi,
\label{eq:pauliEQ}
\end{equation}
but in the case $\sigma_k= (\Id,\Id,\Id)_k$ the spin degrees of freedom decouple from the space degrees and
\eqref{eq:pauliEQ} 
becomes a plain Schr\"odinger equation for a Pauli $N$-particle spinor, equivalent to a system of $2^N$ independent 
original Schr\"odinger equations indexed by the various spin-up-down combinations.
	The current \eqref{eq:jA} then coincides with the Schr\"odinger current of ``spinless electrons'' 
whenever the initial data for $\psi$ are independent of the $\zeta_k$, while \eqref{eq:jB} becomes
the current of electrons with ``constant spins''  \cite{llBOOK}, the last term simplifying to 
$\frac{\hbar}{2m}{\nab_k\left(\psi^\dagger\psi\right)\times s_0},$ where $s_0 = (1,1,1) \forall k$.

	In \eqref{eq:pauliEQ}, $\phi_k = \phi(q_k,t)$ and $A_k = A(q_k,t)$,
with $\phi\in\Rset$ and $A\in \Rset^3$ being electric and magnetic potentials due to \textit{external} 
sources, i.e. $\phi$ and $A$ are not affected by the $N$ electrons. 
	In this ``first quantized'' theory $\phi$ and $A$ are classical potential fields, 
assumed to satisfy the Lorenz-Lorentz gauge condition,
\begin{equation}
\frac{1}{c}\ddt\phi + \nab\cdot A = 0,
\label{eq:LLgauge}
\end{equation}
so that the Maxwell equations for $\phi,A$ become the inhomogeneous wave equations,
\begin{alignat}{1}
        \frac{1}{c^2}\ddtt{\phi({x}, t)}
        - \nab^2 \phi({x}, t)   
&=
        4 \pi \rhoel ({ x}, t) \, ,
\label{eq:MwavePHI}
\\
        \frac{1}{c^2}\ddtt{A({x}, t)}
        - \nab^2 A({x}, t)   
&= 
        4\pi \frac{1}{c}{\jel}({x},t) \, ,
\label{eq:MwaveA}
\end{alignat}
with charge density $\rhoel\in \Rset$ and current density vector $\jel\in\Rset^3$ 
satisfying the continuity equation,
\begin{equation}
	\partial_t \rhoel
	+ \nab\cdot \jel
=
	0
\, ,
\label{eq:continuityLAW}
\end{equation}
which is implied by \eqref{eq:LLgauge}, \eqref{eq:MwavePHI}, \eqref{eq:MwaveA}.
	Otherwise these density functions are prescribed arbitrarily and 
represent electric sources \emph{external} to the system of $N$ electrons, i.e.
$\rhoel=\rhoelext$ and $\jel=\jelext$.

	In all situations where non-electromagnetic effects as well as  electromagnetic effects
due to so-called second quantization are negligible and whenever the electrons move at speeds much 
less than light, this setup works well \emph{for all practical purposes} as long as the  fields generated 
by the external sources are so strong that the contributions from the $N$ electrons to the total 
fields are negligible. 
	The model also explains what is happening rather than having recourse to the obscure 
``measurement formalism.''
	However, even assuming a purely electromagnetic world, the various ``approximations''
are not satisfactory conceptually as long as it is not really clear what consistent model
the above model is approximate to!
	Some of the conceptual inconsistencies are:
\begin{itemize}
\item
 the electrons react passively to electromagnetic fields but do not actively produce fields,
 thus:
  \begin{itemize}
\item
 the external sources evolve in some essentially arbitrarily chosen manner, influencing the electrons 
 through the fields they create without being influenced by the electrons;
  \item
 the electrons neither influence each other nor themselves individually;
\end{itemize}
\item
 the dynamics of the fields, given the sources in one Lorentz frame, satisfies Lorentz-invariant
 equations while the equations for the electrons, given the fields, are Galilei-invariant;
\item
 the electrons satisfy a quantum dynamics, the fields  a classical dynamics.
\end{itemize}
	In this paper we will be chiefly concerned with the first item in the above list, more
precisely with its second subitem.
	Thus, we will not attempt to create a feedback loop from the electrons to the external sources; 
however, it should be clear that such a feedback loop is easily achieved once the problem of how
the $N$ electrons interact mutually is solved consistently --- by simply abandoning the 
somewhat arbitrary split of the ``material'' world into the system of $N$ electrons and 
everything else  (the external sources), treating the whole as a single system.
	Item two, while perhaps the most important one as emphasized by Bell \cite{bellBOOK},
will not be addressed here; yet, see \cite{berndlETal,munchberndlETal, KieJSPa,KieJSPb}.
	We will not address item three either.

	Now, as for the question of how to obtain a consistent interacting theory of electrons and fields, 
the nice thing about taking a de Broglie-Bohm perspective, besides avoiding the measurement
problem, is that an obvious candidate for such a consistent model suggests itself: at the first-quantized level 
discussed here, the electric and magnetic potentials $\phi$ and $A$ in \eqref{eq:pauliEQ} 
should not be considered as created only ``externally'' (not a sharp notion anyhow) but 
instead as being the potentials for the total classical fields.
	Thus, the potentials $\phi$ and $A$ then solve 
\eqref{eq:LLgauge}, \eqref{eq:MwavePHI}, \eqref{eq:MwaveA}, with total charge and current 
densities given by $\rhoel(x,t) = \rhoelext(x,t) - e \sum_{k=1}^N \delta_{q_k(t)}(x)$ and
$\jel(x,t) = \jelext(x,t) - e \sum_{k=1}^N \dot{q}_k(t) \delta_{q_k(t)}(x)$, respectively, 
and with particle velocities $\dot{q}_k$ given by the de Broglie-Bohm guiding law \eqref{eq:dBBguide}. 
	Conceptually this appears to be a consistent setup ---
unfortunately, technically it is not!

	We now have inherited the problems with infinite self-energies and
self-forces that plague(d) classical Lorentz electrodynamics with point charges \cite{ThirringBOOKa}, 
except that the ill-defined total Lorentz forces are replaced by a not 
well-defined r.h.s. of the guiding equation \eqref{eq:dBBguide}.
	Indeed, recall that the solutions of 
\eqref{eq:LLgauge}, \eqref{eq:MwavePHI}, \eqref{eq:MwaveA} for
$\rhoel(x,t) = \rhoelext(x,t) - e \sum_{k=1}^N \delta_{q_k(t)}(x)$ and
$\jel(x,t) = \jelext(x,t) - e \sum_{k=1}^N \dot{q}_k(t) \delta_{q_k(t)}(x)$ are given by a 
linear superposition of Li\'enard-Wiechert solutions 
	       \cite{lienard, wiechert},
fields due to external sources, and a source-free field.
	A Li\'enard-Wiechert solution diverges at the location of the charge;
hence, the r.h.s. in \eqref{eq:jA}, \eqref{eq:jB} are ill-defined.

	At this point, we will follow the classical example of Abraham  \cite{abrahamBOOK} and  
replace the point charges with extended charges.
	Thus we obtain a quantum Abraham model of electrodynamics
(see Appendix A.3 in \cite{AppKieAOP} and \cite{spohnBOOK} for modern discussions of classical Abraham models.)
	The introduction of this model and the discussion of some of its salient features will 
consume the main part of this paper.
	Yet, the main objective of this paper is to convey some interesting
lessons to be learned from contemplating this model.

	I end this introduction with a dedication to GianCarlo Ghirardi, who's quest for a rational 
understanding of quantum physics has led him to develop, jointly with A. Rimini and T. Weber, their
spontaneous localization theory \cite{GRW} which, together with the de Broglie-Bohm theory, is the only
theory of quantum mechanics filed in the reasonable category by the philosopher H.~Putnam \cite{Putnam}.
	With his successful pursuit of objective realistic foundations of physics, GianCarlo 
has been an inspiration to all those scientists, especially the younger ones, who have balked at
the mainstream quantum physics education which maintains that such a pursuit is futile. 
	R.~Tumulka's \cite{RodiGRW} recent construction of the first Lorentz-invariant 
generalization of the Galilei-invariant GRW theory, for not mutually interacting particles, 
speaks testament to my words.
	It gives me great pleasure to celebrate GianCarlo's 70th birthday with this small
contribution that grew out of my own modest attempts to understand the foundations of physics. 

\section{The Quantum Abraham models}

\subsection{Physical setting}
	In order to limit the scope of this article we formulate the models only 
for a neutral atom, treating the nucleus in the Born-Oppenheimer approximation as
an infinitely massive point charge which is fixed at the origin of an inertial frame.
	Hence, the nucleus is treated as an external source, of magnitude $N$ times
the electron |charge|.
	In addition we admit an applied magnetic field, implemented through field conditions
``at infinity.''

\subsection{The kinematical electron according to Abraham}
        As in the pre-relativistic classical works of Abraham~\cite{abrahamBOOK},
a \emph{(bare) charge} of each electron is now distributed around its instantaneous 
location $q_k(t)\in {\mathbb R}^3$ by a compactly supported smooth bare charge density 
$-\eb\fe$ with $SO(3)$ symmetry in {\it all} states of 
motion in the nucleus' inertial frame, $\fe$ satisfying $\int_{{\mathbb R}^3}\fe(x){\dd}^3x = 1$.
        The $k$-th electron carries this charge density along rigidly with velocity 
$\dot{q}_k(t)$; it may also rotate rigidly around $q_k(t)$ with angular velocity ${\omega}_k(t)$.
	In addition the spherical electron will have a \emph{bare mass} $\mb>0$, 
a \emph{bare moment of inertia} $\Ib>0$, and a \emph{bare gyromagnetic ratio} $\gb>0$ 
($\gb =2$ in \eqref{eq:pauliEQ}).

	In classical Abraham theory one can take $\eb=e$, but in the quantum Abraham theory 
we shall see that $\eb\neq e$. 
	Moreover, in the original Abraham model
$\mb=0$ and $\Ib=0$ --- with disastrous consequences; see Appendix A.3 in \cite{AppKieAOP}.

\subsection{The basic equations}
	With charge and current densities $\rhoel$ and $\jel$ thus defined,
the Maxwell wave equations \eqref{eq:MwavePHI}, \eqref{eq:MwaveA} in Lorenz-Lorentz gauge
become semi-relativistic Maxwell-Lorentz wave equations for the electromagnetic potentials 
$\varphi$ and $a$ of the bare charges,
\begin{alignat}{1}
        \frac{1}{c^2}\ddtt{\varphi({x}, t)}
        - \nab^2 \varphi({x}, t)   
&=
        4 \pi \eb\big( N\delta_0(x) -  \sum_{k=1}^N \fe(x-q_k(t))\big)\, ,
\label{eq:MLwavePHI}
\\
        \frac{1}{c^2}\ddtt{a({x}, t)}
        - \nab^2 a({x}, t)   
&= 
 -4\pi \eb\sum_{k=1}^N \fe(x-q_k(t)){\textstyle{\frac{1}{c}}} 
\big(\dot{q}_k(t)+ \omega_k(t)\times \big(x -q_k(t)\big)\big)\, .
\label{eq:MLwaveA}
\end{alignat}
	We remark that the notion of semi-relativistic field equations refers to the failure of the
r.h.s. of \eqref{eq:MLwavePHI} and \eqref{eq:MLwaveA} to transform properly under Lorentz
transformations; indeed, what may appear to be a Minkowski four-vector on the left side in
\begin{equation}
\begin{pmatrix}
 c\rhoel(x,t)\\ 
   \jel(x,t)
  \end{pmatrix}_{\mathrm{electrons}}
 =  
-\eb \sum_{k=1}^N \fe(x-q_k(t) )
\begin{pmatrix}
c\\
\dot{q}_k(t)+ \omega_k(t)\times \big(x -q_k(t)\big)
  \end{pmatrix}
\end{equation}
is clearly not a proper four-vector on the right side, for the rigidity of $\fe$ and the Euler form of the 
rotation velocity $\omega_k(t)\times \big(x -q_k(t)\big)$ do not transform properly.

	The velocity of the $k$-th electron is given by a de Broglie-Bohm guiding law,
either in a version without spin current contribution,
\begin{equation}
 \frac{\dd q_k}{\dd t} 
\Big|_{\qV(t)} = 
\frac{1}{\psi^\dagger\psi}
 {\Re\,\psi^\dagger\left(-i{\textstyle{\frac{\hbar\ }{\mb}}}\nab_k 
	+ {\textstyle{\frac{\eb}{\mb c}}}\fe*_ka\right)\psi}\Big|_{\qV(t)},
\phantom{+ \frac{\gb\hbar}{2\mb}{\nab\times\left(\psi^\dagger\sigma_k\psi\right)}}
\label{eq:dBBguideAL}
\end{equation}
or in a version including it,
\begin{equation}
\quad \frac{\dd q_k}{\dd t} 
\Big|_{\qV(t)} = 
\frac{1}{\psi^\dagger\psi}\left(
{\Re\,\psi^\dagger\left(-i{\textstyle{\frac{\hbar\ }{\mb}}}\nab_k 
	+ {\textstyle{\frac{\eb}{\mb c}}}\fe*_k a\right)\psi}
+{\textstyle{\frac{\gb\hbar}{\mb 4}}}{\nab_k\times\left(\psi^\dagger\sigma_k\psi\right)}\!\right)\!\Big|_{\qV(t)}
\,;
\label{eq:dBBguideALspin}
\end{equation}
in both \eqref{eq:dBBguideAL} and \eqref{eq:dBBguideALspin},
\begin{equation}
\fe *_k a =  \int_{{\mathbb R}^3} \fe(x -q_k) a(x,t)\, \dd^3x\, .
\label{eq:regA}
\end{equation}

	As to the angular velocity of the $k$-th spinning charge, various distinct possibilities exist
too.
	In particular, one can choose either $\omega_k(t)\equiv 0$ (spin is purely ``contextual''), 
or (if spin is at least partly intrinsic) one can define $\omega_k(t)$ through
\begin{equation}
 \omega_k(t)
 = \frac{\gb\hbar}{\Ib 4}
\frac{1}{\psi^\dagger\psi}{\psi^\dagger\sigma_k\psi}\Big|_{\qV=\qV(t)};
\label{eq:omegaEQ}
\end{equation}
yet another possibility will be encountered later.

	The electrons' antisymmetric $N$-particle spinor wave function 
$\psi(\,.\,,t) \in \Cset^{\Rset^{3N}\times\{-1,1\}^N}$ now satisfies the following Pauli equation,
with $\sigma_k= (\sigma^x,\sigma^y,\sigma^z)_k$ or $\sigma_k= (\Id,\Id,\Id)_k$,
\begin{equation}
i\hbar\ddt\psi 
= 
\sum_{k=1}^N \Big({\textstyle{\frac{1}{2\mb}}}
\left(-i\hbar\nab_k + \textstyle{\frac{\eb}{c}}\fe*_k a\right)^2 
-\eb\fe*_k\varphi
+ \gb\textstyle{\frac{\hbar\eb}{4\mb c}}\sigma_k\cdot\fe*_k \beta
\Big)\psi
\,.
\label{eq:regpauliEQ}
\end{equation}
	In \eqref{eq:regpauliEQ}, $\beta = \nab\times a$, and
$\fe*_k \varphi$ and $\fe*_k \beta$ are defined analogously to $\fe*_ka$ in \eqref{eq:regA}.

	The above set of equations comprises several models of quantum Abraham type.
	To the best of my knowledge, none of these models has been discussed in the literature.
	These models open up a new perspective on de Broglie-Bohm
quantum theory, for the following reason. 
	In the conventional completely non-relativistic de Broglie-Bohm theory of, say, 
an atom in an applied magnetic field, the interaction terms are all prescribed and the
Pauli equation for the atomic electrons is a linear equation 
for $\psi$ independent of the evolution of the actual charges. 
	Thus, while $\psi$ guides their motion, there is no feedback loop from the actual motion 
into the dynamics of $\psi$ (which is just fine).
	In the quantum Abraham models, the motion of the actual charges, which is again
guided by the spinor wave function, influences the actual electromagnetic fields whose
potentials enter the Pauli equation. 
	\emph{Thus there is a feedback loop from the motion of the actual particles to the evolution of
the wave function}.\footnote{Lest the reader gets the impression that any Bohmian quantum theory that treats
		the interactions between particles and fields consistently, rather than putting some of them
		in by hand, would have to have a feedback loop from the actual particles to $\psi$, we emphasize
		that a different type of consistent theory is possible in which there is once again no influence
		of the actual particles on $\psi$ \cite{KieJSPa, KieJSPb}.}
	Hence, the  Pauli equation \eqref{eq:regpauliEQ}
for $\psi$ is part of a \emph{nonlinear} system of dynamical equations.
	As a consequence, issues like interference phenomena that seem to rely crucially on having a 
linear evolution of $\psi$ have to be re-evaluated in light of this novel structure. 
	It should therefore be clear that the question, ``which one of the quantum Abraham models, if any, 
reproduces the physical phenomena correctly?'' is an interesting new problem, to which we turn next.

\section{The de Broglie-Bohm version of the pre-standard model of a not too large atom}

\subsection{Outline of the pre-standard model}
	When it comes to computing spectral lines of any atom with $N\geq 1$, but $N$ not too big,
the nucleus assumed fixed, \eqref{eq:pauliEQ} with 
its external potential $\phi$ being the Coulomb potential of the nucleus gives decent
quantitative agreement between experiments and computations only for $N=1$.
	However, if the Coulomb repulsion 
of the electrons, if necessary the Breit-Darwin relativistic correction to it, the spin-orbit and 
spin-spin couplings (see \cite{pauliREVIEW}) are additionally put into the linear Pauli equation 
\eqref{eq:pauliEQ} by hand, a vastly improved quantitative accuracy is achieved.
	We call this amended Pauli equation, coupled with \eqref{eq:dBBguide} and either
\eqref{eq:jA} or \eqref{eq:jB}, the de Broglie-Bohm version of the ``pre-standard model'' 
of a not too large atom, in analogy to its ``standard model'' \cite{froehlichNOTES} which 
treats the electromagnetic radiation field as (second) quantized and which is also known 
as the Pauli-Fierz model or ``non-relativistic QED;'' see \cite{spohnBOOK} for a good survey. 
	Of course, in the conventional (i.e. without de Broglie-Bohm guiding law) pre-standard 
and standard models, contact with physical reality is made through the obscure ``measurement formalism.''
	It ought to be possible to  systematically derive the 
pre-standard model, the measurement formalism included, from a consistent realistic quantum theory.
	In this section we inquire into whether a quantum Abraham theory of electrons and fields 
reduces to the pre-standard model in the so-called adiabatic regime of gentle accelerations
(and low speeds).
	Note that if such a derivation is possible, it then also follows that at least in this adiabatic 
regime the nonlinear evolution of the $\psi$ of the quantum Abraham model will be compatible with interference
effects normally thought to be a consequence of the linearity of the evolution of $\psi$ in the conventional
Schr\"odinger and Pauli equations.

\subsection{The spinless quantum Abraham model}
	To keep matters simple, but also because I know how to do the calculations for the model with spin
only for a system with a \emph{single} electron and a fixed nucleus (the Born-Oppenheimer hydrogen), the
many electrons model will now be considered in the version with $\sigma_k= (\Id,\Id,\Id)_k$, and with 
a guiding equation \eqref{eq:dBBguideAL}.
	Since then the spin indices are reduced to a mere bookkeeping device, we actually ignore them
(factor them out), treating $\psi$ as just the Schr\"odinger wave function, i.e. $\psi\in \Cset^{\Rset^{3N}}$;
of course, $\psi$ is antisymmetric, but the following does not need this input.
	
\subsubsection{The polar representation}
	Inserting the polar representation of the wave function, $\psi = \rho^{1/2}e^{i\Phi^{\mathrm{pol}}}\!\!$,
into the Schr\"odinger equation gives the formally equivalent system of equations \cite{Madelung}
\begin{equation}
\hbar\ddt\Phi 
= 
- \sum_{k=1}^N \Big({\textstyle{\frac{1}{2\mb}}}
\left( \hbar\nab_k \Phi+ \textstyle{\frac{\eb}{c}}\fe*_k a\right)^2 -\eb\fe*_k\varphi\Big) 
+ {\textstyle{\frac{\hbar^2}{2\mb}}}\rho^{-1/2}\vec{\nab}^2 \rho^{1/2}
\,,
\label{eq:regQHJeq}
\end{equation}
\begin{equation}
\ddt\rho
= 
- \sum_{k=1}^N \nab_k\cdot \Big(
\left( {\textstyle{\frac{\hbar\ }{\mb}}}\nab_k\Phi + \textstyle{\frac{\eb}{\mb c}}\fe*_k a\right)\rho\Big)\,,
\label{eq:QcontinuityEQ}
\end{equation}
with potentials $\varphi, a$ still solving \eqref{eq:LLgauge}, \eqref{eq:MLwavePHI}, and
\eqref{eq:MLwaveA} with $\omega_k\equiv 0$.
	In \eqref{eq:regQHJeq} and \eqref{eq:QcontinuityEQ}, $\Phi$ is the ``continuous primitive'' of 
$D\Phi^{\mathrm{pol}}$, a technicality necessary because the definition of $\Phi^{\mathrm{pol}}$ implies
that $\Phi^{\mathrm{pol}}$ jumps by $\pm\pi$ across hypersurfaces $\{\rho^{1/2}=0\}$ of codimension 1 
whenever that node of $\rho^{1/2}$ has odd transversal algebraic multiplicity. 
	The guiding equation \eqref{eq:dBBguideAL} rewrites as
\begin{equation}
 \frac{\dd q_k}{\dd t} 
\Big|_{\qV=\qV(t)} = 
\left({\textstyle{\frac{\hbar\ }{\mb}}}\nab_k\Phi +\textstyle{\frac{\eb}{\mb c}}\fe*_k a\right)\Big|_{\qV=\qV(t)}.
\label{eq:dBBguideALinPHI}
\end{equation}

\subsubsection{The Bohm-type pseudo-classical formulation}
	Next we proceed as if \eqref{eq:regQHJeq}, \eqref{eq:dBBguideALinPHI}
were a Hamilton-Jacobi formulation of the dynamics of classical
charged particles moving under the influence of self-consistently computed electromagnetic fields.
	Of course, the term $\frac{\hbar^2}{2\mb} \rho^{-1/2}\vec{\nab}^2 \rho^{1/2}$ in \eqref{eq:regQHJeq},
the so-called ``quantum potential'' \cite{bohmhileyBOOK}, with $\rho$ satisfying \eqref{eq:QcontinuityEQ},
makes it plain that we are not dealing with classical Hamilton-Jacobi theory, but this term does 
not change the thrust of the argument.
	Namely, taking the time derivative of \eqref{eq:dBBguideALinPHI} we obtain the following
pseudo-Newtonian equation of motion\footnote{While formally
		\eqref{eq:qALeqTmb} and \eqref{eq:pDEFnewton} are just a system of Newton's 
		equations of motion equipped with the electromagnetic Lorentz- and a nonclassical 
		gradient force, in order to reproduce the de Broglie-Bohm trajectories the initial 
		condition $\pkb(0)$ cannot be prescribed independently of $q_k(0)$ and $\psi_0$ 
		but is still determined by the guiding equation \eqref{eq:dBBguideALinPHI}.} 
for the $k$-th electron, 
\begin{equation}
        \frac{\dd \pkb}{\dd t}\Big|_{\qV(t)}
= 
 -\eb \fe*_{k} \left[ {\veps}(\,.\,,t) + {\textstyle{\frac{1}{c}}}\dot{q}_{k}(t) \times{\beta}(\,.\,,t)\right]
\Big|_{\qV(t)}
+  {\textstyle{\frac{\hbar^2}{2\mb}}}\nab_k( \rho^{-1/2}\vec{\nab}^2 \rho^{1/2})\Big|_{\qV(t)}
\label{eq:qALeqTmb}
\end{equation}
where $\pkb$ is the  bare Newtonian momentum of the $k$-th electron,
\begin{equation}
        \frac{\dd q_k}{\dd t}\Big|_{\qV=\qV(t)}
= 
	\frac{1}{\mb}{\pkb}(t)
\,.
\label{eq:pDEFnewton}
\end{equation}
	In \eqref{eq:qALeqTmb}, the classical electric and magnetic fields $\veps ({x},t)\in{\mathbb R}^3$ 
and $\beta({x} ,t)\in{\mathbb R}^3$ at the space point $x\in\Rset^3$ at time $t\in \Rset$ 
are defined through the familiar formulas $\veps = -\nab \varphi - \frac{1}{c}\ddt a$ and 
$\beta = \nab \times a$. 
	The electric and magnetic fields are solutions of the classical 
semi-relativistic Maxwell-Lorentz equation~\cite{abrahamBOOK, lorentzENCYCLOP, lorentzBOOK}, 
which for our atomic setup read
\begin{alignat}{1}
        \nab\cdot{\beta}({ x}, t)  
&= 
        0\, ,
\label{eq:MLdivB}
\\
        \frac{1}{c}\ddt{\beta({x}, t)}
        + \nab\times \veps({x}, t)   
&= 
        {0}\, ,
\label{eq:MLrotE}
\\
        \nab\cdot \veps({ x}, t)  
&=
        4 \pi \eb\big( N\delta_0(x) -  \sum_{k=1}^N \fe(x-q_k(t))\big)\, ,
\label{eq:MLdivE}
\\
        - \frac{1}{c}\ddt{\veps({x},t)}
        + \nab\times{\beta}({ x}, t)  
&= 
 -4\pi\eb\sum_{k=1}^N \fe(x-q_k(t)){\textstyle{\frac{1}{c}}}\dot{q}_k(t)\, ;
\label{eq:MLrotB}
\end{alignat}
note that r.h.s. \eqref{eq:MLrotB} is r.h.s. \eqref{eq:MLwaveA} with $\omega_k\equiv 0$ in 
this spinless model.

\subsubsection{The classical limit}
	We pause for a moment to note that if the term 
$\frac{\hbar^2}{2\mb}\nab_k (\rho^{-1/2}\vec{\nab}^2 \rho^{1/2})$ is dropped from \eqref{eq:qALeqTmb},
the above equations \eqref{eq:qALeqTmb}--\eqref{eq:MLrotB}
do become the spinless Abraham model with bare mass $\mb$ and bare charge $\eb$;
the latter can now be set equal to the empirical charge $e$.
	This vindicates the name quantum Abraham model for our spinless model.

	At this point it is also appropriate to briefly recall a few things that
are known about this spinless  Abraham model with bare mass
(`material mass' in \cite{lorentzENCYCLOP, lorentzBOOK}),
which is the most thoroughly studied simplified model of semi-relativistic Lorentz electrodynamics.
	In particular, we mention the recent rigorous treatments in 
\cite{BauerDuerr, KomechSpohn, KunzeSpohn}, where for technical reasons the 
Newtonian formula \eqref{eq:pDEFnewton} is replaced by Einstein's formula 
\begin{equation}
   \frac{\dd q_k}{\dd t}\Big|_{\qV=\qV(t)}
=
 \frac{1}{\mb}  \frac{\pkb(t)}{\sqrt{1 + |\pkb (t)|^2/(\mb c)^2}}
\,,
\label{eq:pDEFeinstein}
\end{equation}
with $\mb\neq 0$, providing an a priori bound on the particle velocities.
        In this case, of course also the Newtonian kinetic energy ${\textstyle{\frac{1}{2\mb}}}|\pkb |^2$
is to be replaced by Einstein's rest plus kinetic energy $\mb c^2 (1 + |\pkb |^2 /(\mb c)^2)^{1/2}$.
        In \cite{BauerDuerr, KomechSpohn}, the global existence and 
uniqueness for the Cauchy problem of a particle without spin and 
$\mb\neq 0$ was proven. 
        Moreover, in \cite{BauerDuerr} it was shown that the motion
is stable if $\mb >0$ and unstable if $\mb <0$. 
	In \cite{AppKieAOP} it was shown that the ``purely electromagnetic'' models with $\mb=0$
are overdetermined.\footnote{Incidentally, this invalidates the derivation of the  
		Newtonian equation of motion from the purely electromagnetic Abraham model(s) in 
		\cite{abrahamBOOK, lorentzBOOK, panofskyphillipsBOOK, rohrlichBOOK, jacksonBOOK}.
		In \cite{jacksonBOOKrev} a ``more conservative position'' is taken and the spinless 
		Abraham model with bare mass considered.
		There it is also left for the reader to discover (excercise 16.4) that 
		stability of motion requires $\mb>0$; however, the crucial observation that the initial value 
		problem is singular when $\mb=0$ is not made.}
	In \cite{KiePLA} the conservation laws are studied. 

        The paper \cite{KomechSpohn} addresses the long time asymptotics of a spinless 
particle with $\mb >0$, and \cite{KunzeSpohn} studies the adiabatic regime of gentle accelerations. 
	For us the important point is the evaluation of the Lorentz force in this adiabatic regime, 
which is valid irrespectively of how the particles come about to move gently (and, for us, slowly).
	Thus, in the adiabatic regime, the electric and magnetic fields in the vicinity of $q_k(t)$ 
formally split into self fields and non-self fields, viz.
$\veps = \veps^{k} + \veps^{\not\, k}$ and $\beta = \beta^{k} + \beta^{\not\, k}$, 
which are evaluated in a power series in $1/c$.
	To leading order the Lorentz self-force makes a contibution to the inertia of the $k$-th particle, 
i.e. it produces a term $-m_{\mathrm{el}}\ddot{q}_k$ so that the bare inertial mass of the particle becomes 
``renormalized.''
	In leading order for $\veps^{\not\, k}$ one finds the instantaneous Coulomb field of the other
regularized electron bare charges plus the Coulomb field of the point nucleus (the external charge)
with bare charge $N\eb$. 
	In the next order for $\veps^{\not\, k}$ one finds the Lorentz force due to the applied (``external'')
magnetic field, next retarded interactions, and from $\veps^{k}$ radiation-reaction effects, 
see also \cite{spohnBOOK}.

	For later purposes we need the classical mass renormalization formula 
\cite{abrahamBOOK, lorentzBOOK, spohnBOOK},
\begin{equation}
 m_{\mathrm{class}}  = \mb + 
\frac{2\eb^2}{3 c^2} \iint \frac{\fe(x)\fe(y)}{|x-y|}\dd^3{x}\dd^3{y}.
\label{eq:massRENclass}
\end{equation}
	In order to obtain the physically correct Newtonian equations of motion of an 
``atom in the classical regime,'' in the classical spinless Abraham model one would set $\eb = e$ 
and choose $\mb$ such that $m_{\mathrm{class}} = m$, with $-e$ and $m$ the empirical electron charge and mass.

\subsubsection{Derivation of the spinless pre-standard model for short times}
	Radiation-reaction effects are not included in the pre-standard model.
	Low speeds in our model allow us to also neglect retarded interactions among
different particles.

	Abraham's adiabatic approximation applied to the fields in \eqref{eq:qALeqTmb} yields
\begin{alignat}{1}
   \frac{\dd \pkc}{\dd t}\Big|_{\qV=\qV(t)}
= 
&
\left[-
 \sum_{l=1}^N  \eb^2 
\nab_k C(q_k,q_l(t)) + \eb \nab_k \varphi^{\mathrm{ext}}_k
- \eb{\textstyle{\frac{1}{c}}}\dot{q}_k(t) \times{\beta}^{\mathrm{ext}}_k\right]_{\qV=\qV(t)}
\nonumber
\\
&+  {\textstyle{\frac{\hbar^2}{2\mb}}}\nab_k\left(\rho^{-1/2} \vec{\nab}^2 \rho^{1/2}\right)\ \Big|_{\qV=\qV(t)}
\,,
\label{eq:qALeqTmbAD}
\end{alignat}
where 
\begin{equation}
C(q_k,q_l)
\equiv \iint \frac{\fe(x-q_k)\fe(y-q_l)}{|x-y|}\dd^3{x}\dd^3{y}
\end{equation}
and $\nab_k$ acts on the generic $q_k$, not on the actual $q_k(t)$.
	In \eqref{eq:qALeqTmbAD}, $\pkc$ defined by 
\begin{equation}
        \frac{\dd q_k}{\dd t}\Big|_{\qV=\qV(t)}
= 
	\frac{{\pkc}(t)}{m_{\mathrm{class}}}
\label{eq:pDEFnewtonREN}
\end{equation}
is the classically renormalized Newtonian momentum of electron $k$,
with $m_{\mathrm{class}}$ given by \eqref{eq:massRENclass}.
	In deriving \eqref{eq:qALeqTmbAD} we neglected the variation of the external fields over 
$\supp\,\fe$.

	We next fix the bare parameters in this adiabatic regime.
	The quantum potential gradient in \eqref{eq:qALeqTmbAD}
enforces on us a different choice of the bare parameters $\eb$ and $\mb$ 
than in the classical Abraham model, to obtain the correct pre-standard model of an atom from 
our quantum Abraham model.
	After multiplying \eqref{eq:qALeqTmbAD} by $\mb/m$ to get the empirical electron mass
to appear in the quantum potential, and noting that the electromagnetic fields and potentials generated
by the bare charges, i.e. $\veps$ and $\beta$ as well as $\varphi$ and $a$, scale proportional to $\eb$,
we obtain
\begin{alignat}{1}
m_{\mathrm{class}}\mb & = m^2,
\label{eq:massRENq}
\\
\mb \eb^2 & = m e^2,
\label{eq:chargeRENq}
\end{alignat}
with $m_{\mathrm{class}}$ given by \eqref{eq:massRENclass} containing $\mb$ and $\eb^2$.
	The renormalization equations
\eqref{eq:massRENq} and \eqref{eq:chargeRENq} can be solved uniquely for the bare parameters 
$\mb>0$ and $\eb^2 >0$ in terms of the empirical parameters $m$, $e$, and the ``Coulomb energy'' of $\fe$ ---
provided the electrostatic Abraham mass of $-e\fe$ does not exceed the empirical electron mass, 
i.e. provided that
\begin{equation}
\frac{2e^2}{3 c^2} C(0,0)
< m\,.
\label{eq:elMASScondition}
\end{equation}
	Whenever \eqref{eq:elMASScondition} is satisfied, for the bare charge and mass we find,
respectively,
\begin{alignat}{1}
\eb^2 &= e^2 \Big/ \sqrt{1 - \frac{2e^2}{3 m c^2} C(0,0)}\Big.
\,,
\label{eq:ebFIX}
\\
\mb & = m\sqrt{1 - \frac{2e^2}{3 m c^2} C(0,0)}
\,.
\label{eq:mbFIX}
\end{alignat}

\emph{Remarks:}
\begin{itemize}
\item
	Curiously, the quantum renormalization equations \eqref{eq:massRENq} and \eqref{eq:chargeRENq}
do not involve $\hbar$;
\item
	In contrast to the classical renormalization were $\mb>0$ can only be 
inferred from a stability analysis of electron motions \cite{BauerDuerr}, here the positivity
$\mb>0$ already follows from \eqref{eq:chargeRENq};
\item
	Curiously, too, the \emph{solvability condition} \eqref{eq:elMASScondition} 
is precisely the classical condition for $\mb>0$ when
$m_{\mathrm{class}}= m$ with $\eb=e$, although here $m_{\mathrm{class}}\neq m$ and $\eb\neq e$;
\item
	Note that \eqref{eq:elMASScondition} rules out the point charge limit $\fe\to \delta$;
\item
	With ``$<$'' replaced by ``$=$,'' \eqref{eq:elMASScondition} becomes Abraham's formula for the
purely electromagnetic rest mass of a spinless electron.
	It differs from Einstein's $E=mc^2$ by a factor $4/3$ due to the 
model's lack of proper Lorentz invariance, see \cite{Fermi, rohrlichBOOK}.
\end{itemize}

	With the bare parameters given by \eqref{eq:ebFIX} and \eqref{eq:mbFIX},
equations \eqref{eq:qALeqTmbAD} and \eqref{eq:pDEFnewtonREN} become 
\begin{alignat}{1}
   \frac{\dd p_k}{\dd t}\Big|_{\qV=\qV(t)}
= 
&
\left[-
 \sum_{l=1}^N  e^2 
\nab_k C(q_k,q_l(t))
+ e \nab_k \phi^{\mathrm{ext}}_k
- e{\textstyle{\frac{1}{c}}}\dot{q}_k(t) \times{B}^{\mathrm{ext}}_k\right]_{\qV=\qV(t)}
\nonumber
\\
&+  {\textstyle{\frac{\hbar^2}{2m}}}\nab_k\left(\rho^{-1/2} \vec{\nab}^2 \rho^{1/2}\right)\ \Big|_{\qV=\qV(t)}
\,,
\label{eq:qALeqTmAD}
\end{alignat}
where $p_k$ now is the empirical Newtonian momentum of the $k$-th electron, viz.
\begin{equation}
        \frac{\dd q_k}{\dd t}\Big|_{\qV=\qV(t)}
= 
	\frac{{p_k}(t)}{m}
\label{eq:pDEFnewtonEMP}
\end{equation}
with $m$ the empirical electron mass.
	In \eqref{eq:qALeqTmAD} we used
$\frac{\mb}{m}\eb\beta^{\mathrm{ext}} = e B^{\mathrm{ext}}$ and
$\frac{\mb}{m}\eb\varphi^{\mathrm{ext}} = e \phi^{\mathrm{ext}}$,
with $B^{\mathrm{ext}}=\nab \times A^{\mathrm{ext}}$ and $E^{\mathrm{ext}}= - \nab \phi^{\mathrm{ext}}$
solving
\begin{alignat}{1}
        \nab\cdot{B^{\mathrm{ext}}}({ x})  
&= 0
\,,
\label{eq:MLdivBext}
\\
        \nab\times{B^{\mathrm{ext}}}({ x})  
&=  0
\,, 
\label{eq:MLrotBext}
\\
        \nab\cdot E^{\mathrm{ext}}({ x})  
&=
        4 \pi e N\delta_0(x) 
\,,
\label{eq:MLdivEext}
\\
        \nab\times{E^{\mathrm{ext}}}({ x})  
&= 0
\, 
\label{eq:MLrotEext}
\end{alignat}
for a Born-Oppenheimer atom.
	\emph{Given} $\rho^{-1/2}\vec{\nab}^2\rho^{1/2}$, equations 
\eqref{eq:qALeqTmAD}, \eqref{eq:pDEFnewtonEMP} are 
equivalent (locally in $t$) to the following  Hamilton-Jacobi type system
for a phase function $\Phi^{\mathrm|\rho}$ (conditioned on $\rho$) and the actual positions,
where we use that $\phi_k^{\mathrm{ext}} = Ne |q_k|^{-1}$,   
\begin{alignat}{1}
\hbar\ddt\Phi^{\mathrm|\rho}
= 
& - \sum_{k=1}^N \Big({\textstyle{\frac{1}{2m}}}
\Bigl|\hbar\nab_k \Phi^{\mathrm|\rho}+ {\textstyle{\frac{e}{c}}}A_k^{\mathrm{ext}}\Bigr|^2 
-e^2N |q_k|^{-1} \Big) 
- \frac{e^2}{2} \sum_{k=1}^N\sum_{l=1}^N C(q_k,q_l) 
\nonumber
\\
&\qquad\quad + \textstyle{\frac{\hbar^2}{2m}}\rho^{-1/2}\vec{\nab}^2 \rho^{1/2}
\,,\label{eq:regQHJeqCONDrho}
\\
\frac{\dd q_k}{\dd t} 
\Big|_{\qV(t)} 
= &\
\left({\textstyle{\frac{\hbar}{m}}}\nab_k\Phi^{\mathrm|\rho} 
	+ \textstyle{\frac{e}{mc}}A_k^{\mathrm{ext}}\right)\Big|_{\qV=\qV(t)},
\label{eq:dBBguideALinPHIcondRHO}
\end{alignat}
which produces the same dynamics of the actual electrons as does \eqref{eq:regQHJeq}, 
\eqref{eq:dBBguideALinPHI} in the adiabatic regime for $\Phi$, given the same $\rho^{-1/2}\vec{\nab}^2\rho^{1/2}$, 
given the same initial conditions $q_k(0)$, and provided the initial condition $\Phi^{\mathrm|\rho}_0$ 
in the neighborhoods of the $q_k(0)$ is so chosen that the r.h.s.'s of \eqref{eq:dBBguideALinPHI} and 
\eqref{eq:dBBguideALinPHIcondRHO} produce the same actual inital velocities.

	Incidentally,  if once again the quantum potential is dropped, we have just shown that,
in the regime of low speeds and gentle accelerations, 
the \emph{upgraded test particles Hamilton-Jacobi theory} (see \cite{KieJSPa})
of the classical spinless Abraham model with positive bare mass produces the
same actual dynamics as the classical \emph{proper particles Hamilton-Jacobi theory} of 
electrons interacting at a distance and with an external field.

	The remaining issue now is the quantum potential.
	Note that $\rho$ satisfies the continuity equation 
\eqref{eq:QcontinuityEQ} involving $\mb$, $\Phi$, and the $\eb\fe*_k a$ rather than 
$m$, $\Phi^{\mathrm|\rho}$, and the $eA_k^{\mathrm{ext}}$. 
	However, we can choose the initial $\Phi^{\mathrm{|\rho}}_0$ such that
the velocity field which supplies the r.h.s. of \eqref{eq:dBBguideALinPHIcondRHO} 
agrees initially on all of $\Rset^{3N}$ with the one supplying the r.h.s. of \eqref{eq:dBBguideALinPHI}.
	By continuity, for some sufficiently short time span into the future of the initial
instant, these two velocity fields launch essentially identical evolutions of the initial
$\rho$. 
	This in turn implies now that, if they agree initially, the function pair
$(\Phi^{|\rho},\rho)$ and  the solution pair $(\Phi^{\mathrm{alt}},\rho^{\mathrm{alt}})$ 
of the following alternate ``quantum Hamilton-Jacobi PDEs,''
\begin{alignat}{1}
\hbar\ddt\Phi^{\mathrm{alt}}
= 
& - \sum_{k=1}^N \Big({\textstyle{\frac{1}{2m}}}
\left|\hbar\nab_k \Phi^{\mathrm{alt}}+ {\textstyle{\frac{e}{c}}}A_k^{\mathrm{ext}}\right|^2 
-e^2N |q_k|^{-1}\Big) 
- \frac{e^2}{2} \sum_{k=1}^N\sum_{l=1}^N C(q_k,q_l) 
\nonumber
\\
& + \textstyle{\frac{\hbar^2}{2m}}\frac{1}{\sqrt{\rho^{\mathrm{alt}}}}\vec{\nab}^2 \sqrt{\rho^{\mathrm{alt}}}
\,,\label{eq:regQHJeqALT}
\\
\ddt\rho^{\mathrm{alt}}
=& 
- \sum_{k=1}^N \nab_k\cdot \Big(
\left({\textstyle{\frac{\hbar}{m}}}\nab_k\Phi^{\mathrm{alt}} 
	+ \textstyle{\frac{e}{mc}}A_k^{\mathrm{ext}} \right)\rho^{\mathrm{alt}}\Big)
\,,
\label{eq:QcontinuityEQalt}
\end{alignat}
will continue to agree for a comparably short time span, 
and the actual particle motions will then basically agree with the solutions of
\begin{equation}
\frac{\dd q_k}{\dd t} 
\Big|_{\qV(t)} 
= 
\left( {\textstyle{\frac{\hbar}{m}}}\nab_k\Phi^{\mathrm{alt}} 
	+ \textstyle{\frac{e}{mc}}A_k^{\mathrm{ext}}\right)\Big|_{\qV=\qV(t)}
\,.
\label{eq:dBBguideALinPHIalt}
\end{equation}

   Finally, we multiply \eqref{eq:QcontinuityEQalt} by $\pm ie^{i\Phi^{\mathrm{alt}}}/\sqrt{\rho^{\mathrm{alt}}}$
and \eqref{eq:regQHJeqALT} by $\Psi \equiv \pm \sqrt{\rho^{\mathrm{alt}}}e^{i\Phi^{\mathrm{alt}}}$,
the sign chosen on each connected component of the interior of 
$\supp {\rho^{\mathrm{alt}}}$ so as to make $\Psi$ differentiable
across hypersurfaces $\{{\rho^{\mathrm{alt}}}=0\}$ of codimension 1.
	Now adding the so multiplied equations, we find that the wave function $\Psi$
satisfies the Schr\"odinger equation
\begin{equation}
i\hbar\ddt\Psi 
= 
\sum_{k=1}^N
 \Big(
{\textstyle{\frac{1}{2m}}}
 \left(-i\hbar\nab_k + \textstyle{\frac{e}{c}}A_k^{\mathrm{ext}}\right)^2 -e^2N |q_k|^{-1}
+ \frac{e^2}{2} \sum_{l=1}^N C(q_k,q_l) \Big)\Psi
\,.
\label{eq:erwinEQ}
\end{equation}
	In terms of $\Psi$, the guiding equation \eqref{eq:dBBguideALinPHIalt} becomes
\begin{equation}
 \frac{\dd q_k}{\dd t} 
\Big|_{\qV=\qV(t)} = 
{\Re\,\Psi^{-1}\left(-i{\textstyle{\frac{\hbar}{m}}}\nab_k 
	+ {\textstyle{\frac{e}{mc}}}A_k^{\mathrm{ext}}\right)\Psi}\Big|_{\qV=\qV(t)}
\,.
\label{eq:dBBguideERWIN}
\end{equation}
	Equations \eqref{eq:erwinEQ} and \eqref{eq:dBBguideERWIN} constitute the de Broglie-Bohm version 
of the pre-standard model of a not too large spinless atom.
	Although we did not supply analytical estimates, our derivation of this model is rigorous.
	The above derivation is valid for sufficiently short times.
	In fact, the solution $\Psi\cong (\Phi^{\mathrm{alt}},\rho^{\mathrm{alt}})$ 
of \eqref{eq:erwinEQ} will in general agree with $(\Phi^{|\rho},\rho)$  \emph{only} for short times.

\subsubsection{Empirical equivalence over longer times?}

	Because we invoked continuity qualitatively, we cannot quantify how short ``sufficiently short'' really is.
	However, we will argue (nonrigorously) that, while the pairwise agreement on $\Rset^{3N}$ 
of $\Psi\cong (\Phi^{\mathrm{alt}},\rho^{\mathrm{alt}})$ with $(\Phi^{|\rho},\rho)$ is most likely restricted 
to times shorter than any characteristic atomic time scale, the actual electron motions as computed
from \eqref{eq:erwinEQ}, \eqref{eq:dBBguideERWIN} and those computed from 
\eqref{eq:regQHJeq}, \eqref{eq:QcontinuityEQ}, \eqref{eq:dBBguideALinPHI} in the adiabatic regime 
should nevertheless agree well over much longer times!

	As for the various wave functions, the problem is that, even though we know that \emph{given} 
$\rho^{-1/2}\vec{\nab}^2\rho^{1/2}$ computed from \eqref{eq:regQHJeq}, \eqref{eq:QcontinuityEQ} 
the guiding equations \eqref{eq:dBBguideALinPHI} and \eqref{eq:dBBguideALinPHIcondRHO} produce (essentially) 
the same \emph{actual} electron motions in the adiabatic regime (also for longer times), the velocity fields 
on $\Rset^{3N}$ which supply the r.h.s.'s of \eqref{eq:dBBguideALinPHI} and \eqref{eq:dBBguideALinPHIcondRHO}
will generally \emph{not agree} on all of $\Rset^{3N}$ at later times \emph{even if they agree initially}.
	The reason is simply that for the (quantum Abraham) Hamilton-Jacobi equation \eqref{eq:regQHJeq}
almost all generic points in $N$ particle configuration space represent test (i.e. passive)
particles in an environment produced by the $N$ actual particles (and some external sources and the 
quantum potential (ES-QP)), while for the more conventional Hamilton-Jacobi type equation \eqref{eq:regQHJeqCONDrho}
each generic configuration point represents $N$ active generic particles interacting with each other 
electrostatically (and otherwise with the same ES-QP)
	In other words, in the setup \eqref{eq:regQHJeqCONDrho}, any single generic particle experiences
the influence of $N-1$ other typically generic but active particles plus the ES-QP,
while in the setup  \eqref{eq:regQHJeq} any single generic particle typically experiences the influence of $N$ 
other particles which are the actual ones, plus the ES-QP, the exception occurring
precisely when the $N$ generic positions coincide with the $N$ actual ones --- in which case
the test particles are ``upgraded'' to active particles.
	This significant difference in the velocity fields on the generic configuration points
means that not long after the initial instant of time the densities 
$\rho$ and $\rho^{\mathrm{alt}}$ will begin to differ.
	It follows that the velocity field on $\Rset^{3N}$ which supplies the r.h.s. of 
\eqref{eq:dBBguideALinPHIcondRHO} will soon differ from the one supplying the r.h.s. 
of \eqref{eq:dBBguideALinPHIalt} even though they agree initially, too.
	Thus, at those times the velocity field represented by the 
r.h.s. of \eqref{eq:dBBguideERWIN} will begin to differ from the ones
represented by \eqref{eq:dBBguideALinPHIcondRHO} and by \eqref{eq:dBBguideALinPHI}.
	An easy rule of thumb estimate is that in a typical generic atomic configuration significant
differences will develop in  a fraction of a typical orbital time in an atom.

	And yet the actual motions as computed from \eqref{eq:erwinEQ}, \eqref{eq:dBBguideERWIN} 
should nevertheless continue to agree with those computed from \eqref{eq:regQHJeq}, \eqref{eq:QcontinuityEQ}, 
\eqref{eq:dBBguideALinPHI} in the adiabatic regime for much longer times.
	To vindicate this extraordinary claim we note first that from \eqref{eq:qALeqTmAD} and the
corresponding rewriting of \eqref{eq:dBBguideERWIN} it follows that, in order to produce the
same actual motions, $\rho^{-1/2}\vec{\nab}^2\rho^{1/2}$ and 
$\sqrt{\rho^{\mathrm{alt}}}^{-1}\vec{\nab}^2\sqrt{\rho^{\mathrm{alt}}}$ need only agree along those
trajectories. 
	The crucial observation now is that, despite the generic differences of the velocity
fields on the generic configuration space computed with the upgraded test particles theory versus
the proper (interacting) particles theory, the smooth regularization of the point charges implies
that \emph{in the immediate vicinity} of an actual position point of an atomic configuration, a test 
particle experiences predominantly the electric field of the $N-1$ other actual particles (plus the ES-QP), 
i.e. the situation resembles what happens in the proper particles theory with configurations
that basically agree with the actual ones. 
	This conclusion is true for typical atomic configurations when the mutual interparticle distance
is about a Bohr radius (or even only a tenth of it) while the size of the regularizing $\fe$ is smaller than that
by as much as a factor $\alpha^2$ in the spinless model, where $\alpha$ is Sommerfeld's fine structure constant. 
	Of course, there is also the contribution from $\rho$ versus the one from $\rho^{\mathrm{alt}}$.
	But, a bootstrap argument now reveals that, if initially the two densities $\rho$ and 
$\rho^{\mathrm{alt}}$ agree and so do the velocity fields computed with \eqref{eq:erwinEQ}, 
\eqref{eq:dBBguideERWIN} versus those computed from \eqref{eq:regQHJeq}, \eqref{eq:QcontinuityEQ}, 
\eqref{eq:dBBguideALinPHI} in the adiabatic regime, then in configuration space there will exist 
a thin tubular region  around the actual trajectory computed from  \eqref{eq:regQHJeq}, 
\eqref{eq:QcontinuityEQ}, \eqref{eq:dBBguideALinPHI} in the adiabatic regime,
which should have a cross section size which is a fraction of the diameter of the support of $\fe$, 
in which the velocity field corresponding to \eqref{eq:dBBguideALinPHI} and the density 
$\rho$ continue to agree, respectively, with the velocity field corresponding to 
\eqref{eq:dBBguideALinPHIalt} and the density $\rho^{\mathrm{alt}}$ for a considerably longer time than
away from this tubular region.

	\emph{Hence, as to its empirial output of actual particle motions, the spinless quantum Abraham model 
in the adiabatic regime should make the same predictions as the de Broglie-Bohm version of the spinless 
pre-standard model of a not too large atom over times much longer than an atomic period.}
	What the de Broglie-Bohm version of the spinless pre-standard model achieves is a 
``parallel processing of all possibly actual motions which can be computed with the quantum Abraham model,
launched by an ensemble of possible initial data,''
\emph{the} actual motion being selected \emph{a posteriori} by the actually actual initial data, 
while the quantum Abraham model requires the actual particle initial data (and field initial 
data, if we are not in the adiabatic regime) \emph{a priori}, producing 
an ensemble of test particle motions but only \emph{one actual motion ``at a time''}.

	\subsubsection{Stationary Coulomb states}

	As is well known, stationary states of \eqref{eq:erwinEQ} are of the form 
$\Psi(\qV,t)= e^{-\hbar^{-1}Et} \Xi(\qV)$, with $\Xi(\qV)$ solving the stationary
Schr\"odinger equation.
	Setting now the applied magnetic field to zero, i.e. $A^{\mathrm{ext}}=0$, the stationary
Schr\"odinger equation for $\Xi$ becomes
\begin{equation}
\sum_{k=1}^N
 \Big(
-{{\frac{\hbar^2}{2m}}}\Delta_k - Ne^2 \frac{1}{|q_k|}
+ \frac{e^2}{2} \sum_{l=1}^N  \iint \frac{\fe(x - q_k)\fe(y-q_l)}{|x-y|}\dd^3{x}\dd^3{y}\Big)\Xi
= 
E \Xi 
\label{eq:erwinEQstat}
\end{equation}
	The energy differences computed from the eigenvalue problem \eqref{eq:erwinEQstat} agree well
with the empirical spectral data if fine details due to the finite nuclear mass, electron spin and 
relativity are ignored. 
	The velocity field represented by the r.h.s. of \eqref{eq:dBBguideERWIN} becomes
\begin{equation}
\vV =
 {\Re\,\Xi^{-1}\left(-i{\textstyle{\frac{\hbar}{m}}}\nab_k \right)\Xi} = \vec{0}
\label{eq:dBBguideERWINstat}
\end{equation}
for $\Xi$ can be chosen real; hence, the actual particles are at rest by \eqref{eq:dBBguideERWIN}.
	Since, as we have argued in the previous subsection, the empirical output regarding the actual
particle motions is the same in the quantum Abraham model and the de Broglie-Bohm version of the
pre-standard model, the actual particles are also at rest according to  
\eqref{eq:regQHJeq}, \eqref{eq:QcontinuityEQ}, \eqref{eq:dBBguideALinPHI}, 
and we are then surely in the adiabatic regime. 
	The question is whether the spectral lines computed from  \eqref{eq:regQHJeq}, \eqref{eq:QcontinuityEQ}
with the actual configuration agree with those of the pre-standard model. 
	Setting $a=0$ also in \eqref{eq:regQHJeq}, and setting furthermore $\Phi = - \hbar^{-1}{\cal E} t$, 
which then also implies through \eqref{eq:dBBguideALinPHI} that the actual particles are at rest, 
\eqref{eq:regQHJeq} becomes the following stationary Schr\"odinger equation for $\Upsilon\equiv \pm\rho^{1/2}$, 
\begin{equation}
\sum_{k=1}^N
 \Big(
-{{\frac{\hbar^2}{2m}}}\Delta_k - Ne^2 \frac{1}{|q_k|}
+ e^2 \sum_{l=1}^N  \iint \frac{\fe(x - q_k)\fe(y-q_l[0])}{|x-y|}\dd^3{x}\dd^3{y}\Big)\Upsilon
= 
{\cal E} \Upsilon,
\label{eq:erwinEQstatABRAHAM}
\end{equation}
where the $\!q_l[0]\!$ are the equilibrium positions of the $N$ electrons. 
	Clearly, the two Schr\"odinger potentials in \eqref{eq:erwinEQstat} and \eqref{eq:erwinEQstatABRAHAM}
are quite different, but the equilibrium positions in \eqref{eq:erwinEQstatABRAHAM} are to be found from 
equating the r.h.s. of \eqref{eq:qALeqTmb} to $0$ (with $\dot{q}_l = 0$). 
	Whether  \eqref{eq:erwinEQstat} and \eqref{eq:erwinEQstatABRAHAM}
are iso-spectral (after at most a translation of $E$ or $\cal E$) is an interesting new problem.
	In particular, the hydrogen problem becomes an exactly solvable two-center Schr\"odinger
problem coupled with a force balance  problem for equilibrium positions of the electron 
which we hope to address in some future work. 
 (See comments added in the appendix.)

\subsection{Quantum Abraham model with spin of Born-Oppenheimer hydrogen in an applied magnetic field}

	\subsubsection{Choice of the quantum Abraham model}
	A priori there is still some freedom in choosing the model.
	In this subsection we discuss equations 
\eqref{eq:LLgauge}, \eqref{eq:MLwavePHI}, \eqref{eq:MLwaveA}, \eqref{eq:regpauliEQ}  for $N=1$, 
together with \eqref{eq:dBBguideAL} and \eqref{eq:omegaEQ}.
	We proceed in the same spirit as for the spinless model.

	\subsubsection{The Cayley-Klein representation}
	In terms of Euler angles $(2\Phi,\Theta,\Omega)$, the Cayley-Klein representation of $\psi$ reads 
\begin{equation}
\psi = \pm \rho^{1/2} e^{i\Phi} \begin{pmatrix}
				\cos (\Theta/2) e^{i\Omega/2} \\
				i\sin(\Theta/2) e^{-i\Omega/2}
				\end{pmatrix}
\,,
\label{eq:CKrepOFpsi}
\end{equation}
where the meaning of the ``$\pm$'' is the same as before.
	Inserting \eqref{eq:CKrepOFpsi} into \eqref{eq:regpauliEQ} for $N=1$ yields an
equivalent system of first order equations (cf., \cite{Takabayasi, BST, BS, S, Reginatto, PGU}),
comprising the following ``quantum Hamilton-Jacobi PDE''
\begin{alignat}{1}
\hbar\left(\ddt\Phi +  \half\cos\Theta \ddt\Omega\right)
= 
& - {\textstyle{\frac{1}{2\mb}}}
\Bigl( \hbar (\nabE \Phi+ \half\cos \Theta \nabE \Omega) 
	+ {\textstyle{\frac{\eb}{c}}}\fe*_{\mathrm{e}} a\Bigr)^2 
\nonumber\\ 
& + \eb\fe*_{\mathrm{e}}\varphi
- \gb{\textstyle{\frac{\eb}{2 \mb c}}} s \cdot (\fe*_{\mathrm{e}}\beta)
\label{eq:regQHJeqBST}
\\
& - {\textstyle{\frac{\hbar^2}{8\mb}}}\left((\nabE \Theta)^2 + (\sin\Theta)^2( \nabE \Omega)^2\right) 
	+ {\textstyle{\frac{\hbar^2}{2\mb}}}\rho^{-1/2}\Delta_{\mathrm{e}} \rho^{1/2} 
\nonumber\,,
\end{alignat}
where $_{\mathrm{e}}$ indicates the electron's generic configuration space variable and 
\begin{equation}
s = {{\frac{\hbar}{2}}}  \begin{pmatrix}
\sin(\Theta)\sin(\Omega)\\
\sin(\Theta)\cos(\Omega)\\
\cos(\Theta)
				\end{pmatrix}
\label{eq:SPINfield}
\end{equation}
is the spin vector field of $\psi$
(i.e. $s={{\frac{\hbar}{2}}}{\psi^\dagger\sigma\psi}/{\psi^\dagger\psi}$ evaluated with \eqref{eq:CKrepOFpsi});
the following continuity equation for the density,
\begin{equation}
\ddt\rho
= 
-  \nabE\cdot \Big({\textstyle{\frac{1}{\mb}}}
\Bigl(\hbar (\nabE \Phi+ \half\cos \Theta \nabE \Omega) 
		+ {\textstyle{\frac{\eb}{c}}}\fe*_{\mathrm{e}} a\Bigr)\rho\Big)
\,;
\label{eq:QcontinuityEQbst}
\end{equation}
the following two transport equations for the spin field, 
\begin{alignat}{1}
\ddt \cos\Theta + 
{\textstyle{\frac{1}{\mb}}}
\Bigl(\hbar (\nabE \Phi+ \half\cos \Theta \nabE \Omega) 
		+ {\textstyle{\frac{\eb}{c}}}\fe*_{\mathrm{e}} a\Bigr)
\cdot\nabE\cos\Theta 
& =
\ \frac{2}{\hbar}\frac{\delta H_{\mathrm{spin}}}{\delta \Omega}
\label{eq:THETAeqBST}
\\
\ddt \Omega + 
{\textstyle{\frac{1}{\mb}}}
\Bigl(\hbar  (\nabE \Phi+ \half\cos \Theta \nabE \Omega) 
		+ {\textstyle{\frac{\eb}{c}}}\fe*_{\mathrm{e}} a\Bigr)
\cdot\nabE\Omega 
& =
- \frac{2}{\hbar}\frac{\delta H_{\mathrm{spin}}}{\delta \cos\Theta}
\,,
\label{eq:XIeqBST}
\end{alignat}
where
\begin{equation}
H_{\mathrm{spin}} = \int \rho\left(
{\textstyle{\frac{\hbar^2}{8\mb}}}\left((\nabE \Theta)^2 + (\sin\Theta)^2( \nabE \Omega)^2\right) 
+ \gb {\textstyle{\frac{\eb}{2\mb c}}} s \cdot (\fe*_{\mathrm{e}}\beta)\right)\dd{q_{\mathrm{e}}}.
\label{eq:BSTspinHAMILTONIAN}
\end{equation}
	The electromagnetic potentials $\varphi, a$ solve 
\eqref{eq:LLgauge}, \eqref{eq:MLwavePHI}, \eqref{eq:MLwaveA} for $N=1$, with the actual motion 
of the electron\footnote{The theory of a point electron moving according to the 
		de Broglie-Bohm guiding equation for a Pauli equation with \emph{external} electromagnetic fields 
		(hence, needing no regularization with $\fe$) is developed in \cite{bohmhileyBOOK};
		such an electron motion is a ``test particle'' motion.
		Curiously enough, in \cite{BST} not the de Broglie-Bohm interpretation is used
		but instead a fluid-mechanical interpretation of the Pauli equation, developed
		independently in \cite{Takabayasi} in the spirit of Madelung's early work on the 
		one-particle Sch\"odinger equation \cite{Madelung}.
		Thus, in the BST work $\rhoel=-e\psi^\dagger\psi$ is the real charge density and 
		$\jel = -e {\Re\,\psi^\dagger\left(-i{\textstyle{\frac{\hbar\ }{m}}}\nab_{\mathrm{e}}
			+ {\textstyle{\frac{e}{m c}}}A_{\mathrm{e}}\right)\psi}$
		the real electric current vector density of the electron fluid. 
		Since in their work there are then no true particles, the electromagnetic potentials are not
		convoluted with $\fe$; strangely, BST treat them as given, although logically the model
		should be, and could easily be completed by computing the potentials self-consistently from  
		\eqref{eq:LLgauge} and \eqref{eq:MLwavePHI}, \eqref{eq:MLwaveA} with the above $\rhoel$ 
		and $\jel$ as sources.
		The so completed BST model would simply become what would go under the name \emph{classical
		Maxwell-Pauli equations} (cf., the \emph{classical Maxwell-Dirac equations} \cite{flatoetal}).
		This continuum analog of the  quantum Abraham equations studied in this paper 
		is not known to describe the physics correctly; yet
		``second quantizing'' these ``classical Maxwell-Pauli equations'' yields 
		``non-relativistic QED'' \cite{froehlichNOTES} --- formally, for these
		quantum field equations are UV and IR divergent without cutoffs.
		Of course, the fluid ontology of the classical Maxwell-Pauli equations disappears in this
		heuristic process of ``second quantization.''}
determined by the guiding equation
\begin{equation}
 \frac{\dd q_{\mathrm{e}}}{\dd t} \Big|_{q_{\mathrm{e}}=q_{\mathrm{e}}(t)} 
= 
{\textstyle{\frac{1}{\mb}}}
\Bigl(\hbar (\nabE \Phi+ \half\cos \Theta \nabE \Omega) 
		+ {\textstyle{\frac{\eb}{c}}}\fe*_{\mathrm{e}} a\Bigr)
\Big|_{q_{\mathrm{e}}=q_{\mathrm{e}}(t)},
\label{eq:BSTguideAL}
\end{equation}
and it's angular velocity  $\omE(t)$ by\footnote{Yet another possibility to define $\omE$ is
			through the dynamics of the Euler angles at $q_{\mathrm{e}}=q_{\mathrm{e}}(t)$.}
\eqref{eq:omegaEQ}, which in view of \eqref{eq:SPINfield} now reads
\begin{equation}
 \sbE(t)\equiv \Ib \omE(t)
 = 
{\textstyle{\frac{\gb}{2}}}s \big|_{q_{\mathrm{e}}=q_{\mathrm{e}}(t)} \,,
\label{eq:omegaEQsb}
\end{equation}
where we also {defined} the bare spin $\sbE$ of the spherical electron. 

\subsubsection{The pseudo-classical formalism}
	We again proceed as if we were to derive the Newtonian formalism from a classical Hamilton-Jacobi 
theory and now derive the pseudo-classical formalism for the quantum Abraham model.

	The time derivative of \eqref{eq:BSTguideAL} yields, after a straightforward calculation and
integration by parts to exchange $\fe*_{\text{e}}$ with the gradient, a pseudo-Newtonian equation of 
motion,
\begin{alignat}{1}
       & \frac{\dd \pb}{\dd t}\Big|_{\qE(t)}
= 
 -\eb \fe*_{\text{e}} \left[ {\veps}(\,.\,,t) 
+ {\textstyle{\frac{1}{c}}}\dot{q}_{\text{e}}(t) \times{\beta}(\,.\,,t)
+ {\textstyle{\frac{1}{\mb c}}} \nab \left(\beta(\,.\,,t)\cdot \sbE(t)\right)\right]\Big|_{\qE(t)}
\nonumber\\ 
&\ \ \ 
-  {\textstyle{\frac{\hbar^2}{2\mb}}}\left(
 {\textstyle{\frac{1}{4\rho}}}\nabE\cdot \left( \rho
\left( \nabE\Theta\otimes\nabE\Theta + \sin^2\theta \nabE\Omega\otimes\nabE\Omega \right)\right) 
 -
\nabE\Big( {\textstyle{\frac{1}{\sqrt{\rho}}}}\nabE^2 \sqrt{\rho}\Big)\right)\Big|_{\qE(t)}
\label{eq:qBSTeqTbare}
\end{alignat}
(cf. \cite{BST,PGU}), 
with the bare Newtonian momentum $\pb$ already defined in \eqref{eq:pDEFnewton}.
	The time derivative of \eqref{eq:omegaEQsb} yields a pseudo-Euler equation of evolution 
(cf. \cite{BST,PGU}),
\begin{equation}
        \frac{\dd \sbE}{\dd t}\Big|_{\qE(t)}
= 
 \sbE(t) \times \left(- \gb{\textstyle{\frac{\eb}{2\mb c}}}\fe*_{\text{e}}\beta(\,.\,,t)
+ {\textstyle{\frac{1}{\mb\rho}}} \nabE\cdot \left( \rho \nabE s \right)\right) \Big|_{\qE(t)}
\,.
\label{eq:qBSTeqRbare}
\end{equation}

The electric and magnetic fields $\veps (\,.\,,t)\in{\mathbb R}^3$ 
and $\beta(\,.\,,t)\in{\mathbb R}^3$ in \eqref{eq:qBSTeqTbare}, \eqref{eq:qBSTeqRbare} 
satisfy the classical semi-relativistic Maxwell-Lorentz 
equation for Born-Oppenheimer hydrogen,
\begin{alignat}{1}
        \nab\cdot{\beta}({ x}, t)  
&= 
        0\, ,
\label{eq:MLdivBhyd}
\\
        \frac{1}{c}\ddt{\beta({x}, t)}
        + \nab\times \veps({x}, t)   
&= 
        {0}\, ,
\label{eq:MLrotEhyd}
\\
        \nab\cdot \veps({ x}, t)  
&=
        4 \pi \eb\big(\delta_0(x) -  \fe(x-\qE(t))\big)\, ,
\label{eq:MLdivEhyd}
\\
        - \frac{1}{c}\ddt{\veps({x},t)}
        + \nab\times{\beta}({ x}, t)  
&= 
 -4\pi\eb \fe(x-\qE(t)){\textstyle{\frac{1}{c}}}\left(\dot{q}_{\text{e}}(t)
+ \omE(t)\times \big(x -\qE(t)\right)\, .
\label{eq:MLrotBhyd}
\end{alignat}

\subsubsection{The classical limit}
	Upon dropping the last line in \eqref{eq:regQHJeqBST} and the term $\propto\hbar^2$ in 
\eqref{eq:BSTspinHAMILTONIAN}, the Cayley-Klein representation of our quantum Abraham model with spin 
becomes a classical Hamilton-Jacobi theory of a spinning charge coupled to the total electromagnetic fields;
cf. \cite{S} for such a classical theory with ``external fields'' only.
	Its Newtonian re-formulation consists of \eqref{eq:qBSTeqTbare}-\eqref{eq:qBSTeqRbare} 
without the terms featuring $\rho$, coupled with  \eqref{eq:MLdivBhyd}-\eqref{eq:MLrotBhyd}.
	This classical system is \emph{not} manifestly identical to the classical equations of an
Abraham electron with spin and bare inertias (and modified torque to account for $\gb$), which couple 
\eqref{eq:MLdivBhyd}-\eqref{eq:MLrotBhyd} with
\begin{equation}
        \frac{\dd\pb}{\dd t}\Big|_{\qE(t)}
= 
        -\eb\int_{{\mathbb R}^3}
        \left[{\veps}({x},t) + {\textstyle{\frac{1}{c}}}v_{\text{e}}({x},t) \times{\beta}({x},t)\right]
        \fe({x} -\qE(t)) \, \dd^3x
\label{eq:ALeqTmb}
\end{equation}
\begin{equation}
        \frac{\dd{s}_{\text{b}}}{\dd t}\Big|_{\qE(t)}
= 
       -\eb {\textstyle{\frac{\gb}{2}}} \int_{{\mathbb R}^3} \big({x} - \qE(t)\big)\times
        \left[\veps({x},t) + {\textstyle{\frac{1}{c}}} v_{\text{e}}({x},t)  \times\beta({x},t)\right]
        \fe({ x} -\qE(t)) \, \dd^3x\, ,
\label{eq:ALeqRmb}
\end{equation}
where $\pb$ is defined in \eqref{eq:pDEFnewton}, $\sbE$ in the first identity in \eqref{eq:omegaEQsb},
and where
\begin{equation}
v_{\text{e}}({x},t) 
= \dot{q}_{\text{e}}(t)+ \omE(t)\times \big(x -\qE(t)\big)
\, .
\label{eq:ALvfield}
\end{equation}
	Yet, if we split $\veps$ and $\beta$ into the electron's self fields 
$\veps^{\text{e}}$ and $\beta^{\text{e}}$ and the
applied fields $\veps^{\not\,\text{e}}$ and $\beta^{\not\,\text{e}}$, we can 
show that the applied contributions to the two 
classical formulations agree in excellent approximation, and that the self contributions correspond 
to mass and spin renormalizations (which may differ).

	Since the applied field varies little over the support of $\fe$, it suffices to
prove that the applied field parts of the two classical formulations agree in leading non-vanishing order 
w.r.t. a Taylor expansion of $\beta^{\not\,\text{e}}(x,t)$ about $x=\qE(t)$.
	For Newton's equation of motion we need to show that 
\begin{equation}
{\textstyle{\frac{1}{\mb}}}  
\fe*_{\text{e}} \left(
\nabE\!\bigl(\beta^{\not\,\text{e}}(\qE,t)\cdot \sbE(t)\bigr)\right)\!\!\Big|_{\qE(t)}
=\!
\int_{{\mathbb R}^3}\!\!\big(\omE(t)\times 
\big(x -\qE(t)\big)\big)\times\beta^{\not\,\text{e}}({x},t) \fe({x} -\qE(t)) \dd^3x
\label{eq:DIPOLEforce}
\end{equation}
in leading order.
	Inserting 
$\beta^{\not\,\text{e}}(x,t) = 
	\beta^{\not\,\text{e}}(\qE(t),t) + (x-\qE(t))\cdot\nabE\beta^{\not\,\text{e}}(\qE,t)|_{\qE(t)} +\dots$
into both sides of \eqref{eq:DIPOLEforce} and recalling that $\fe$ is spherically symmetric, we find that the zero 
order term contributes to the l.h.s. of \eqref{eq:DIPOLEforce} but not
the linear term, while to the r.h.s. of \eqref{eq:DIPOLEforce} the zero order term does not contribute
but the linear term does.
	Evaluated to leading non-vanishing order in each side, \eqref{eq:DIPOLEforce} yields
\begin{equation}
-{\textstyle{\frac{\eb}{\mb c}}}  
\nabE \left(\beta^{\not\,\text{e}}(\qE,t)\cdot \sbE(t)\right)\big|_{\qE(t)}
=
\nabE \left(\beta^{\not\,\text{e}}(\qE,t)\cdot \mub (t)\right)\big|_{\qE(t)}
\,,
\label{eq:DIPOLEforceAPPLIED}
\end{equation}
where $\mub (t)= -\frac{\eb}{3c}\int_{{\mathbb R}^3} |x|^2 \fe({x}) \, \dd^3x\, \omE(t)$ is the 
\emph{bare magnetic moment} of the electron.
	Equation \eqref{eq:DIPOLEforceAPPLIED} displays on both sides the correct coupling of 
$\omE$ to the gradient of the applied field, and also features the familiar relation between 
spin and magnetic moment with correct $g$ factor 2 --- here of course w.r.t. the bare quantities. 
	Similarly, for Euler's equation of motion we need to show that 
\begin{equation}
{\textstyle{\frac{1}{\mb}}}  \sbE(t) \times 
\big(\fe*_{\text{e}}\beta^{\not\,\text{e}}(\,.\,,t)\big)\big|_{\qE(t)}
=\!
 \int_{{\mathbb R}^3}\! \big({x} - \qE(t)\big)\times
        \left[ v_{\text{e}}({x},t)\times\beta^{\not\,\text{e}}({x},t)\right]
        \fe({ x} -\qE(t)) \dd^3x 
\label{eq:DIPOLEtorqueB}
\end{equation}
in leading order w.r.t. the Taylor series for $\beta^{\not\,\text{e}}({x},t)$, and furthermore that 
\begin{equation}
 \int_{{\mathbb R}^3} \big({x} - \qE(t)\big)\times
        \veps^{\not\,\text{e}}({x},t) \fe({ x} -\qE(t)) \, \dd^3x = 0.
\label{eq:DIPOLEtorqueE}
\end{equation}
	Beginning with \eqref{eq:DIPOLEtorqueE}, we note that $\veps^{\not\,\text{e}}(x,t) = -\eb\nab |x|^{-1}$
is the nucleus' Coulomb potential.
	Since, $x\times \nab |x|^{-1} =0$, the l.h.s. of \eqref{eq:DIPOLEtorqueE} is proportional
to $\qE(t)\times\int (\nab |x|^{-1})\fe({ x} -\qE(t)) \, \dd^3x$,  but
$\int (\nab |x|^{-1})\fe({ x} -\qE(t)) \, \dd^3x \propto \qE(t)$ by the spherical symmetry of $\fe(\,.\,)$,
proving  \eqref{eq:DIPOLEtorqueE}.
	Turning thus to \eqref{eq:DIPOLEtorqueB}, we insert the Taylor expansion of $\beta^{\not\,\text{e}}({x},t)$ 
and find that the zero order term contributes to the left and right sides (on the right to the integral 
involving $\omE$ but not to the one involving  $\dot{q}_{\text{e}}$).
	Comparing the leading non-vanishing contributions, \eqref{eq:DIPOLEtorqueB} thus becomes 
\begin{equation}
- {\textstyle{\frac{\eb}{\mb c}}} \sbE(t) \times \beta^{\not\,\text{e}}(\qE(t),t)
=
\mub(t) \times \beta^{\not\,\text{e}}(\qE(t),t)
\label{eq:DIPOLEtorqueBlead}
\end{equation}
in perfect harmony with \eqref{eq:DIPOLEforceAPPLIED}. 
	We remark that the linear Taylor term contributes to the integrals on the right side
 of  \eqref{eq:DIPOLEtorqueB} involving $\omE$ and $\dot{q}_{\text{e}}$, but not to the left side,
so that discrepancies show up proportional to the gradient of  $\beta^{\not\,\text{e}}({x},t)$. 
	We also remark that \eqref{eq:DIPOLEtorqueE} will generally be false with more complicated
applied electric fields.

	Coming now to the self-fields, they do vary appreciably over the support of $\fe$ and we 
cannot employ the above Taylor series expansion to these fields.
	Instead we need to invoke the expansion of the Li\'enard-Wiechert solutions in a Taylor series 
of the retarded time about the current instant $t$, see \cite{spohnBOOK, jacksonBOOK}.
	For the Abraham model with spin that was done in great detail in \cite{Schwarzschild}, and
for the classical limit of the quantum Abraham model with spin the computation can be done accordingly.
	In either case one can show that the first non-vanishing contribution comes with the second order 
time derivative of the retarded time. 
	However, the contribution is quite complicated in each setting and we refrain from elaborating on 
it in any detail. 
	We note that part of the complication is due to the semi-relativistic setting of the models; see
\cite{AppKieAOP} and \cite{spohnBOOK} for computing self-contributions in the relativistic Lorentz model.
        In any event, the main point is that in both models there now arises a magnetic contribution 
to the electron's mass and to its moment of inertia. 

	\subsubsection{The adiabatic regime and the pre-standard model of hydrogen}
	Proceeding as for the spinless model, we now find that in the adiabatic regime 
\begin{alignat}{1}
 &  \frac{\dd p_{\text{bf}}}{\dd t}\Big|_{\qE(t)}
= 
 -\eb \left[- \eb \nabE |\qE|^{-1}
+ {\textstyle{\frac{1}{c}}}\dot{q}_{\text{e}}(t) \times{\beta}^{\not\,\text{e}}(\qE,t)
+ {\textstyle{\frac{1}{\mb c}}} \nabE \left(\beta^{\not\,\text{e}}(\qE,t)\cdot \sbE(t)\right)\!\right]\!
_{\qE=\qE(t)}
\nonumber\\ 
&\
-  {\textstyle{\frac{\hbar^2}{2\mb}}}\left(
 {\textstyle{\frac{1}{4\rho}}}\nabE\cdot \left( \rho
\left( \nabE\Theta\otimes\nabE\Theta + \sin^2\theta \nabE\Omega\otimes\nabE\Omega \right)\right) 
 -
\nabE\Big( {\textstyle{\frac{1}{\sqrt{\rho}}}}\nabE^2 \sqrt{\rho}\Big)\right)\Big|_{\qE(t)}
\label{eq:qBSTeqTbareAD}
\end{alignat}
where $p_{\text{bf}}$ is the quasi-classically renormalized Newtonian momentum of the electron, 
\begin{equation}
        \frac{\dd \qE}{\dd t}\Big|_{\qV=\qV(t)}
= 
	\frac{1}{\mbrot + \mfrot}{p_{\text{bf}}}(t)
\,,
\label{eq:pDEFnewtonBSTren}
\end{equation}
where in leading order $\mbrot +\mfrot = m_{\mathrm{class}} +$ a quadratic function in $\omega$
involving $\Ib$ and $\eb$, where $\omega$ is the frequency in a stationary renormalized electron 
(with $m_{\mathrm{class}}$ given in \eqref{eq:massRENclass}); moreover, 
the electromagnetic field mass $\mfrot$ is a quadratic functional of $\fe$
(cf. \cite{AppKieAOP, AppKieLMP} for the classical Lorentz model).
	Similarly, \eqref{eq:qBSTeqRbare} becomes 
\begin{equation}
        \frac{\dd \sbf}{\dd t}\Big|_{\qE(t)}
= 
- \gb{\textstyle{\frac{\eb}{2\mb c}}} \sbE(t) \times \beta^{\not\,\text{e}}(\qE(t),t)
+  \sbE(t) \times {\textstyle{\frac{1}{\mb\rho}}} \nabE\cdot \left( \rho \nabE s \right) \Big|_{\qE(t)}
\label{eq:qBSTeqRbareAD}
\end{equation}
where $\sbf = (\Ibrot + \Ifrot)\omE$ is the sum of $\sbE(t)$ (see \eqref{eq:omegaEQsb}) and a field spin,
with $\Ifrot\propto\eb^2$. 

	Next one multiplies \eqref{eq:qBSTeqTbareAD} and \eqref{eq:qBSTeqRbareAD} by $\mb/m$ 
to get the empirical electron mass $m$ rather than $\mb$ to appear in the generalized quantum 
potential terms, and demands that the remaining coefficients all take their empirical values.
	Recalling that the electromagnetic fields  $\veps$, $\beta$ and potentials $\varphi$, $a$ 
generated by the bare charges scale proportional to $\eb$,
we now obtain the following system of renormalization equations
for the bare parameters,

\begin{alignat}{1}
(\mbrot+\mfrot)\mb &= m^2,
\label{eq:massRENqSPIN}
\\
\mb \eb^2 & = m e^2,
\label{eq:chargeRENqSPIN}
\\
\gb \eb^2 &= 2 e^2, 
\label{eq:gRENqSPIN}
\\
\mb (\Ibrot +\Ifrot)/\Ib &= m
\,,
\label{eq:inmomRENqSPIN}
\end{alignat}
which again implies that $\mb>0$. 
	There is still some ambiguity here, which can be removed by demanding
that the bare mass is distributed by a positive density $\fm=\fe$,
in which case $\Ib\propto\mb$.
	Rather than entering a detailed discussion of \eqref{eq:massRENqSPIN}-\eqref{eq:inmomRENqSPIN},
assuming instead that the remaining bare parameters $\mb,\eb,\gb$ and $\omega$ can be solved for 
(involving $\fe$) in terms of $m,e,c,\hbar$, so renormalized the
equations of motion become
\begin{alignat}{1}
 &       \frac{\dd p\emp}{\dd t}\Big|_{\qE(t)}
= 
 -e \left[- e \nabE |\qE|^{-1}
+ {\textstyle{\frac{1}{c}}}\dot{q}_{\text{e}}(t) \times{B}^{\not\,\text{e}}(\qE,t)
+ {\textstyle{\frac{1}{m c}}} \nabE \left(B^{\not\,\text{e}}(\qE,t)\cdot s\emp(t)\right)\right]
\Big|_{\qE=\qE(t)}
\nonumber\\ 
&\quad
-  {\textstyle{\frac{\hbar^2}{2m}}}\left[
 {\textstyle{\frac{1}{4\rho}}}\nabE\cdot \left( \rho
\left( \nabE\Theta\otimes\nabE\Theta + \sin^2\theta \nabE\Omega\otimes\nabE\Omega \right)\right) 
 -
\nabE\Big( {\textstyle{\frac{1}{\sqrt{\rho}}}}\nabE^2 \sqrt{\rho}\Big)\right]_{\qE(t)},
\label{eq:qBSTeqTbareADemp}
\\
  &      \frac{\dd s\emp}{\dd t}\Big|_{\qE(t)}
= 
- {\textstyle{\frac{e}{m c}}} s\emp(t) \times B^{\not\,\text{e}}(\qE(t),t)
+  s\emp(t) \times {\textstyle{\frac{1}{m\rho}}} \nabE\cdot \left( \rho \nabE s \right) \Big|_{\qE(t)},
\label{eq:qBSTeqRbareADemp}
\end{alignat}
where $p\emp$ is the empirical Newtonian momentum of the electron, i.e.
\begin{equation}
        \frac{\dd \qE}{\dd t}\Big|_{\qV=\qV(t)}
= 
	\frac{1}{m}{p\emp}(t)
\,,
\label{eq:pDEFnewtonBSTemp}
\end{equation}
and where $s\emp$ is the empirical spin of the electron, i.e.
\begin{equation}
 s\emp(t)
 = 
 s \big|_{q_{\mathrm{e}}=q_{\mathrm{e}}(t)} .
\label{eq:omegaEQsEMP}
\end{equation}

	The remaining steps parallel our discussion of the spinless model.
	Clearly, \eqref{eq:qBSTeqTbareADemp} and \eqref{eq:qBSTeqRbareADemp}
obtain from a phase function $\Phi^{|\rho,s}$ conditioned on the fields $\rho$ and $s$ 
being given, which satisfies 
\begin{alignat}{1}
\hbar\!\left(\ddt\Phi^{|\rho,s} +  \half\cos\Theta \ddt\Omega\right)\!
= 
& - {\textstyle{\frac{\hbar^2}{2m}}}
\Bigl( \nabE \Phi^{|\rho,s}+ \half\cos \Theta \nabE \Omega 
	+ {\textstyle{\frac{e}{\hbar c}}}A_{\text{e}}^{\not\,\text{e}}\Bigr)^2 
\! + {\textstyle{\frac{e^2}{|\qE|}}} - {\textstyle{\frac{e}{m c}}} s \cdot B_{\mathrm{e}}^{\not\,\text{e}}
\nonumber\\
& - {\textstyle{\frac{\hbar^2}{8m}}}\left((\nabE \Theta)^2 + (\sin\Theta)^2( \nabE \Omega)^2\right) 
	+ {\textstyle{\frac{\hbar^2}{2m}}}\rho^{-1/2}\Delta_{\mathrm{e}} \rho^{1/2} 
\,,
\label{eq:regQHJeqBSTcond}
\end{alignat}
with $\rho(\qE,t)$ and $s(\qE,t)$ still satisfying 
\eqref{eq:QcontinuityEQbst}, \eqref{eq:THETAeqBST}, \eqref{eq:XIeqBST}. 
	With properly matched initial conditions, (only) for times shorter than characteristic atomic times
the triple $(\Phi^{|\rho,s},\rho,s)$ will agree on \emph{all} of configuration space with a triple 
$(\Phi^{\mathrm{alt}},\rho^{\mathrm{alt}},s^{\mathrm{alt}})\cong \Psi$ satisfying
\begin{equation}
i\hbar\ddt\Psi 
= 
 \Big({\textstyle{\frac{1}{2m}}}
 \left( \sigma\cdot\left(-i\hbar\nabE + \textstyle{\frac{e}{c}}A_{\mathrm{e}}^{\not\,\text{e}}\right)\right)^2 
-e^2 |\qE|^{-1}\Big)\Psi,
\label{eq:pauliEQhydro}
\end{equation}
and \eqref{eq:pDEFnewtonBSTemp} becomes
\begin{equation}
 \frac{\dd \qE}{\dd t} \Big|_{\qE=\qE(t)} 
= 
\frac{1}{\Psi^\dagger\Psi}{\Re\,\Psi^\dagger\left(-i{\textstyle{\frac{\hbar}{m}}}\nabE
	+ {\textstyle{\frac{e}{mc}}}A_{\text{e}}^{\not\,\text{e}}\right)\Psi}\Big|_{\qE=\qE(t)}
\label{eq:dBBguidePAULI}
\end{equation}
which, by a similar bootstrap reasoning as for the spinless model, 
should agree with \eqref{eq:qBSTeqTbareADemp} over times considerably longer than 
the characteristic atomic times.
	This suggests that the quantum Abraham model with spin in the adiabatic regime 
and  \eqref{eq:pauliEQhydro}, \eqref{eq:dBBguidePAULI} should be empirically equivalent.

\section{Closing remarks}
	While certainly not proven rigorously, the results presented in this paper
suggest quite strongly the empirical equivalence of the pre-standard model of not too large
atoms with some nonlinear quantum Abraham model in which there is a feedback loop from
the actual electron motions to the wave equation. 
	This is somewhat surprising at first, but if proven rigorously (if that is
possible) it would shed an interesting new light on the possible meanings of wave functions.
	The wave function in a quantum Abraham model governs an ensemble indexed by
initial positions, not of possible actual particles as in the (pre-)standard model,
but of test particles --- one and only one configuration of which coinciding with 
the actual particle configuration.
	Yet in terms of \emph{the} actual dynamics its active guiding role is the same as
in the conventional setting. 
	
	Another interesting lesson learned is that
contrary to widespread folklore it seems that one \emph{can} picture an electron
as a ``charged spinning miniature billiard ball'' quite well in  non-relativistic quantum theory,
at least for the hydrogenic problem.
	The two main reasons usually offered for why this feat should be impossible are:
(i) Lorentz' classical calculation that such an electron would have to spin
with about 10 times the speed of light at its equator to produce a magnetic moment of 
the size of the Bohr magneton, and (ii) the ``classically indescribable two-valuedness'' of spin,
see \cite{TomonagaBOOK}.
	However, as to (i), Lorentz' calculation assumed $\mb=0$ and $\Ib=0$
at the outset, which was shown in \cite{AppKieAOP} to be inconsistent
--- with $\mb>0$ and $\Ib>0$ there is no superluminal rotation speed; and as to (ii), the de Broglie-Bohm
perspective teaches us that the ``two-valuedness'' is a matter of the law of motion, not
any intrinsic particle property.

	I end with the disclaimer that I am \emph{not at all} suggesting that the electron were a 
spinning miniature billiard ball. 
	In fact, I believe  \cite{KieJSPa, KieJSPb} that there are very good reasons to try to
construct a theory with point electrons, among them absence of empirical evidence to the 
contrary and logical simplicity. 
	In this vein the old ideas of Abraham and Lorentz, when combined with those of 
de Broglie and Bohm, go a long way teaching us what kind of theoretical models are possible 
at all. 
	In particular, aiming for a theory with point electrons, one might want to think of $\fe$ 
just as a mathematical regularizer rather than representing a physical quality of the electron, but 
as such it should be removed in a limit $\fe\to \delta$. 
	However, precisely this is not possible because there is no such 
thing as negative bare mass for a quantum Abraham model.
	To make a consistent theory with point electrons and fields one has to try something else, 
as done in \cite{KieJSPa, KieJSPb}.
	

\bigskip
\textbf{ACKNOWLEDGEMENTS}:
 Work supported by NSF grant DMS 0406951.

\section*{Appendix (added in March 2007)}
 After the original paper was published I noticed a few things that
deserve being mentioned in this added appendix.
\bigskip

1) One typo, which has now been corrected in this version, occurred
a total of four times in the section 
``The spinless quantum Abraham model:''
namely, in the discussion of renormalization equations in the subsections 
``The classical limit'' 
and 
``Derivation of the spinless pre-standard model for short times,''
a factor $1/6\pi$, conspicuous in front of the bilinear form of 
$\fe$, viz. $C(0,0)$, had to be replaced by $2/3$ in \eqref{eq:massRENclass},
\eqref{eq:elMASScondition}, \eqref{eq:ebFIX}, \eqref{eq:mbFIX}.
 The factor $1/6\pi$ would result if the source terms in the Maxwell--Lorentz
equations were written without the factor $4\pi$, which is a different choice 
of 
electromagnetic units occasionally used in the literature on electromagnetism 
(e.g. in \cite{spohnBOOK}) alternately to the Gaussian units I chose (which 
are used, e.g. in \cite{jacksonBOOK}).

\smallskip

2) The question I raised in the last subsection of the section
``The spinless quantum Abraham model,'' 
namely whether the two different stationary Schr\"odinger equations
\eqref{eq:erwinEQstat} and \eqref{eq:erwinEQstatABRAHAM}
could perhaps be iso-spectral (up to an overall shift) for 
equilibrium positions in \eqref{eq:erwinEQstatABRAHAM} found from 
equating the r.h.s. of \eqref{eq:qALeqTmb} to $0$ (with $\dot{q}_l = 0$, 
for all $l$), can easily be answered in the negative \emph{in general}.
   Namely, setting r.h.s. \eqref{eq:qALeqTmb} to $0$ (with $\dot{q}_l = 0$, 
for all $l$) does not impose any new restriction on the equilibrium positions 
$q_l[0]$ beyond the already known $\dot{q}_l = 0$, so that any choice of 
$\{q_l[0]\}$, $l=1,...,N$, will
be a legitimate set of equilibrium positions (as in conventional Bohmian 
mechanics). 
   Now consider the special case $N=1$ (the hydrogen atom). 
   While iso-spectrality of \eqref{eq:erwinEQstatABRAHAM} and
\eqref{eq:erwinEQstat} obviously obtains in the limit $q_1[0]\to\infty$ 
(but which clearly is not a desirable ``location'' for $q_1[0]$), in the limit 
$q_1[0]\to 0$ \eqref{eq:erwinEQstatABRAHAM} has no eigenvalues at all! 
   The latter follows right away from the Cwickel--Lieb--Rosenbljum 
inequality for the number $N_0$ of bound states of a single-particle 
Schr\"odinger operator, which for \eqref{eq:erwinEQstatABRAHAM} with 
$q_1[0]\to 0$ yields
\begin{equation}
N_0 \leq \frac{8}{3\sqrt{\pi}} \left(\frac{2me^2}{\hbar^2} R\right)^{3/2}
\end{equation}
where $R$ is the radius of the support of $\fe$; for a non-spinning electron 
$R\approx \alpha \lambda_C$, where $\lambda_C$ is the Compton wavelength of 
the electron and $\alpha= \frac{e^2}{\hbar c}$ is Sommerfeld's fine structure 
constant. 
 Thus we find in good approximation
$N_0 \leq \frac{8}{3\sqrt{\pi}} (2\alpha^2)^{3/2} \ll 1$,
hence no bound states.
 Since the spectrum varies continuously with $q_1[0]$, we conclude that in 
general the two eigenvalue problems are not isospectral.
\smallskip

3) The renormalization calculations with spin follow the earlier work of Appel
and the author, and Spohn, in which works 
the inertia moment is computed for the choice $\fe=\fm$, ``for simplicity,''
where $\mb\fm$ is the bare mass density of the particle.
   However, allowing $\fm\neq\fe$ not only seems to be more ``natural,''
it does result in a more satisfactory set of renormalization equations.
   To go through the whole calculation accordingly would blow the frame 
of this appendix; I plan to treat the problem elsewhere. 



\bibliographystyle{aipproc}   

\end{document}